\documentclass[12pt]{iopart}
\pdfoutput=1
\usepackage{amsfonts}
\usepackage{graphicx}

\usepackage{xcolor}
\usepackage{subfigure}
\usepackage{textcomp}
\usepackage{cite}
\usepackage{float}
\usepackage{soul}
\usepackage{stackrel}
\usepackage{hyperref}

\definecolor{dgreen}{rgb}{0,0.7,0}

\expandafter\let\csname equation*\endcsname\relax
\expandafter\let\csname endequation*\endcsname\relax
\usepackage{amsmath,amssymb}

\pdfoutput=1
\usepackage{esint}

\usepackage{color}
\definecolor{dgreen}{rgb}{0,0.7,0}

\begin{document}

\title[]{First-passage Brownian functionals with stochastic resetting}

\author{Prashant Singh$^{1}$ and Arnab Pal$^{2}$}

\address{$^{1}$International Centre for Theoretical Sciences, Tata Institute of Fundamental
Research, Bengaluru 560089, India}
\address{$^{2}$The Institute of Mathematical Sciences, CIT Campus, Taramani, Chennai 600113, India \& Homi Bhabha National Institute, Training School Complex, Anushakti Nagar, Mumbai 400094, India}
\ead{prashant.singh@icts.res.in;~arnabpal@imsc.res.in}
\vspace{10pt}

\begin{abstract}
 We study the statistical properties of first-passage time functionals of a one dimensional Brownian motion in the presence of stochastic resetting.  A first-passage functional is defined as $V=\int_0^{t_f} Z[x(\tau)]$ where $t_f$ is the first-passage time of a reset Brownian process $x(\tau)$, i.e., the first time the process
crosses zero. In here, the particle is reset to $x_R>0$ at a constant rate $r$ starting from $x_0>0$ and we 
focus on the following functionals: (i) local time $T_{loc} = \int _0^{t_f}d \tau ~ \delta (x-x_R)$, (ii) residence time $T_{res} = \int _0^{t_f} d \tau ~\theta (x-x_R)$, and (iii) functionals of the form $A_n = \int _{0}^{t_f} d \tau [x(\tau)]^n $ with $n >-2$. For first two functionals, we analytically derive the exact expressions for the moments and distributions. Interestingly, the residence time moments reach minima at some optimal resetting rates. A similar phenomena is also observed for the moments of the functional $A_n$. Finally, we show that the distribution of $A_n$ for large $A_n$ decays exponentially as $\sim \text{exp}\left( -A_n/a_n\right)$ for all values of $n$ and the corresponding decay length $a_n$ is also estimated. In particular, exact distribution for the first passage time under resetting (which corresponds to the $n=0$ case) is derived and shown to be exponential at large time limit in accordance with the generic observation. This behavioural drift from the underlying process can be understood as a ramification due to the resetting mechanism which curtails the undesired long Brownian first passage trajectories and leads to an accelerated completion.
We confirm our results to high precision by numerical simulations.

\end{abstract}

\section{Introduction}
\label{introduction}
Brownian functionals appear ubiquitously in many different disciplines spanning across physics, stochastic processes, finance, computer science and mathematics (see \cite{Majumdar2005}
for a comprehensive review of this topic). The celebrated 
Feynman-Kac formalism has been instrumental to understand the 
statistical properties of the functionals of a one-dimensional Brownian
motion by translating the classical diffusion problems to the quantum mechanics \cite{Kac1949}. Study of such Brownian functionals also appears in stochastic thermodyanmics in the form of fluctuation theorems which express universal properties of the statistical distribution for functionals like work, heat or entropy change, evaluated along the
fluctuating trajectories taken from ensembles with some well specified initial distributions \cite{Seifert,Jarzynski}. 
In this paper, we focus on another class of functionals that has also attracted quite a lot of attention, namely the ``first-passage Brownian
functionals''. More precisely, let us consider a diffusive particle in one dimension whose position $x(\tau)$ is governed by the following equation of motion
\begin{align}
\frac{dx}{d\tau} = \eta(\tau),
\label{BM-lanevin}
\end{align}
where $\eta(\tau)$ is the Gaussian white noise with zero mean and correlation $\langle \eta(\tau) \eta(\tau') \rangle =2 D \delta (\tau-\tau')$. Here, $D$ is the diffusion coefficient. The particle is initially located at $x_0>0$. Moreover, we consider that there is an absorbing boundary at the origin and let 
$t_f$ denote the first-passage time of the Brownian particle to the origin starting from position $x_0$. Clearly $t_f$ is a random variable and fluctuates between realizations. A functional constructed along such a trajectory 
\begin{align}
V = \int _{0}^{t_f} Z[x(\tau)] d \tau~,
\label{fun-eq-1}
\end{align}
is thus random and defined as a first passage functional  \cite{Majumdar2005}. Here, $Z(x)$ can, in principle, be an arbitrary function of $x$. Given this form, one is usually interested in the various statistical properties of the functional $V$.

A well-studied Brownian functional in the literature is the local time for which $Z(x) = \delta(x-x_{\ell})$. It measures the total time that the particle spends in the vicinity of a desired location $x_{\ell}$ in space  \cite{Majumdar2005,Levy,Knight,Feller,graph,Yor,SinghKundu2021}. As such, the local time provides interesting information with
regard to the spatio-temporal properties of the particle’s trajectory and is thus a useful measure with various applications in chemical reactions and catalytic processes \cite{Agmon1,Agmon2,porousD,reactive}. Another example of such functional is the residence or occupation time which characterizes the total time that the particle spends in a given region of 
space \cite{Majumdar-Comtet2002}. For instance, the celebrated ``arcsine law'' of L\'evy describes the probability distribution of the time spent by a one dimensional Brownian particle on the positive
side of the origin out of the total given time $t$ \cite{Majumdar2005,Levy,Redner2001}. For this class of functionals, $Z(x) = \theta(x-x_{\ell})$, where $\theta(x-x_{\ell})$ denotes the Heaviside function.
In the past, the residence time has been studied
in various different scenarios such as diffusion in confinement \cite{Grebenkov2007}, or in a potential landscape \cite{Sabhapandit:2006}, in heterogeneous diffusion processes \cite{SinghHDP}, in Brownian excursion processes \cite{Louchard} and active models \cite{Singharc}. While these studies have mostly considered residence time for fixed time, we here focus on a situation where the residence time is estimated upto the first passage event. Another generic functional that has been extensively studied is of the form $Z(x) = x^n$ where $n$ is some real number. This observable has been studied both in the context of fixed time \cite{Majumdar2005} or random time ensemble \cite{Kearney2005,Meerson2020}. 
In particular, for $n=0$ in Eq. \eqref{fun-eq-1}, $V$ simply represents the first-passage time $t_f$ which has a myriad of applications in all fields (see \cite{Redner2001,Bray2013} for pedagogic introduction and applications of first passage processes). Similarly for $n=1$  in Eq. \eqref{fun-eq-1}, $V$ represents the total area swept by the diffusing particle till its first-passage time. The study of this area has found applications in sandpile models, percolation models and queueing theory \cite{Kearney2005,Kearney2004,Prellberg1995, Majumdar2007, Dhar1989}. In compact directed percolation on square lattice, the area of the staircase polygon is related to the $n=1$ case \cite{Kearney2005,Prellberg1995} whereas in queuing theory this area is related to the cumulative waiting-time experienced by the customers during busy period \cite{Kearney2004}. 
The first-passage area has also been studied in a one-dimensional jump-diffusion process \cite{FPA-jump}, drifted Brownian motion \cite{FPA-dd}, L\'evy process \cite{FPA-levy} and Ornstein-Uhlenbeck process \cite{FPA-OU-1,FPA-OU-2}. The case $n=-\frac{3}{2}$ was shown to represent the lifetime of a comet within the ambit of random walk theory \cite{Hammersley1961}. The case $n=-\frac{1}{2}$ is related to the period of oscillation of a particle in disordered systems modeled by Sinai potential \cite{Dean2001}. We refer to \cite{Majumdar2005} for a review on the applications of these ``first passage Brownian functionals'' for the \textit{reset-free processes}. In this work, our central goal is to extend our understanding of first passage Brownian functionals in the presence of stochastic resetting \cite{Evansrev2020,Restart1,Restart2}. As we will see, resetting has significant effects on the dynamics which results in non-trivial and distinct changes in the statistics of first-passage time observables.

Stochastic resetting is a renewal process where the dynamics is repeated after some random or fixed time. Although very simple to describe, this mechanism is in fact quite natural to many processes around us. For example, unbinding events in a chemical reaction \cite{ReuveniEnzyme1,Restart-Biophysics3}, cleavage in RNA polymerization \cite{bio-1} or dissociation kinetics of GTP-RhoA in cell contraction \cite{bio-2} can be understood as resetting events. The phenomena has been catalyzed even further since over the last decade, resetting has found overreaching applications in statistical physics \cite{Restart3,Restart4,Restart5,Restart6,PalJphysA,local-r,occup-r-1,occup-r-2,ss,TUR}, computer science \cite{Luby,algorithm}, ecology \cite{SP-0,HRS,Montanari}, complex systems \cite{nonlinear-1,volta,pop} operation research \cite{OR-1,OR-2,OR-3} and economics \cite{eco-1,eco-2,eco-3}. Recently, the field has also seen advancements in experiments \cite{expt-1,expt-2,Faisamt2021}. A paradigm model in the field is the Brownian motion with stochastic resetting for which many interesting results exist \cite{PalJphysA,restart_conc4,restart_conc5,Restart12,Restart13,Restart17, Restart18,Restart19}. We refer to \cite{Evansrev2020} for a review on the state of the art of the subject and \cite{inspection} for a perspective
on its connection with the inspection paradox. In particular, the latter pinpoints to how/when stochastic resetting expedites completion of arbitrary stochastic processes.

The essential idea that stochastic trajectories governed by resetting dynamics share is the following: Consider a particle whose position $x(t)$ evolves according to Eq. \eqref{BM-lanevin} starting from $x(0)=x_0~(>0)$. Motion of the particle is then stopped intermittently at a rate $r$ and it is instantaneously brought to a position $x_R~(>0)$. Following the resetting event, the particle starts diffusing again until the next resetting event occurs. The microscopic evolution equation for the particle can then be written as
\begin{align}
x(t+\Delta t)=
\begin{cases}
&  x(t) + \eta (t) \Delta t, ~~~~~~\text{with prob } (1-r \Delta t), \\
&  x_R,  ~~~~~~~~~~~~~~~~~~~~\text{with prob } r \Delta t.
\end{cases}
\label{update}
\end{align}
where recall that $\eta(t)$ is the Gaussian white noise with zero mean and variance $\langle \eta (t) \eta(t') \rangle =2 D \delta(t-t')$. For simplicity, we set $D=\frac{1}{2}$ without any loss of generality. Moreover, we will assume that there is an absorbing boundary at the origin, and the process ends as soon as the particle hits the boundary. As in the reset-free process, we will denote the first passage time under resetting also by $t_f$. Our aim is to investigate the statistical properties of the functional in Eq. \eqref{fun-eq-1} for the motion that is governed by Eq. \eqref{update}. It is important to note that numerous studies have been made on $t_f$ unraveling many key features: expedition of the first-passage time process perhaps being the most remarkable one \cite{Evansrev2020,Restart1, Restart2,Restart7,Restart8,Restart11,branching}. For example, it is known that the moments of area functional for simple diffusion ($r=0$ case) are infinite \cite{Kearney2005,Meerson2020}. This is a natural consequence of the fact that the first passage time density for the simple diffusion has a power law tail at large time \cite{Redner2001}. In contrast, first passage time density of a reset Brownian process falls exponentially at large times \cite{Evansrev2020,local-r}. So, one would expect that the area will attain finite moments under the resetting mechanism. These crucial observations serve as a motivation for us to extend the Feynman-Kac framework to the first passage Brownian functionals in the presence of resetting. In particular, we provide a comprehensive analysis of the following functionals:
\begin{enumerate}
\item \textit{Local time}: The first functional that we consider is the local time $T_{loc}$ which refers to the amount of time (density) spent in the neighbourhood of position $x_{\ell}$ till its first-passage time. For this case, $Z(x) = \delta(x-x_{\ell})$ and thus $T_{loc}(x_\ell)=\int_0^{t_f}~d\tau~\delta(x(\tau)-x_\ell)$. In here, we will set $x_{\ell} = x_R$ so that we would be measuring the local time near the resetting location.
\item \textit{Residence time}: The second observable goes under the name occupation/residence time for which $Z(x) = \theta (x-x_R)$ with $\theta (y)$ being the Heaviside theta function. Thus, the residence time takes the form $T_{res}(x_R)=\int_0^{t_f}~d\tau~\theta(x(\tau)-x_R)$ and estimates the cumulative time spent by the particle in the region $x > x_R$ till its absorption at the origin.
\item Finally, we look at a class of functionals with $Z(x) = x^n~\text{for }n >-2$. We denote them by $A_n(x_0) = \int _{0}^{t_f} d \tau [x(\tau)]^n$. We keep $x_0$ in the argument to indicate the initial position. Our focus would be to understand the consequences of resetting on the distribution and moments of $A_n(x_0)$.
\end{enumerate}

For the convenience of the readers, we provide a short outline of the paper in the following. In Sec. \ref{backward}, we derive a backward differential equation for the moment generating function of $V(x_0)$. Deploying this backward equation, we then study local time in Sec. \ref{local}, residence time in Sec. \ref{residence} and functional $A_n(x_0)$ in Sec. \ref{area}. Finally, we conclude with some future outlook in Sec. \ref{conclusion}.

\section{General formulation}
\label{backward}
In this section, we show how to compute the probability distribution function (PDF) $P_R(V,x_0)$ of a Brownian functional $V$ over the time interval $[0,t_f]$ where $t_f$ is the first-passage time of
the process in the presence of stochastic resetting.
Our starting point would be to
derive a differential equation for the moment generating function defined as follows
\begin{align}
Q(p,x_0) &=\int _{0}^{\infty} dV ~e^{-p V}~P_R(V,x_0), \\
&=\langle e^{-p \int _{0}^{t_f} Z[x(\tau)] d \tau} \rangle,
\label{def-Q}
\end{align}
where the average $\langle .. \rangle$ in the second line denotes averaging over all trajectories as well as over $t_f$. This enables us to compute the moments of $V$ as
\begin{align}
\langle V^m \rangle = (-1)^m \left(\frac{\partial^m Q(p,x_0)}{\partial p^m}\right) \bigg|_{p \to 0}.
\label{moms-Q}
\end{align} 
We would like to construct now a backward differential equation for $Q(p,x_0)$. 
To proceed, consider a typical trajectory $\{ x(\tau);0 \leq \tau \leq t_f \}$ that starts from $x_0>0$ at $\tau =0$ and split into two parts: (i) a left interval $[0,\Delta t]$ and (ii) a right interval $[\Delta t,t_f]$ with $\Delta t \to 0$. In the first interval, the position of the particle changes from $x_0$ to $x_0'=x_0 + \Delta x$ (where the exact form of $\Delta x$ will be specified later). Starting from this new position $x_0'$ at time $\Delta t$, the particle reaches the origin at time $t_f$. This path-decomposition leads us to break the integration in Eq. \eqref{def-Q} as $\int _{0}^{t_f} = \int _{0}^{\Delta t}+\int _{\Delta t}^{t_f}$ and in the limit $\Delta t \to 0$, we get
\begin{align}
Q(p,x_0) &= \langle e^{-p Z(x_0) \Delta t} ~e^{-p \int _{\Delta t}^{t_f} Z[x(\tau)] d \tau} \rangle, \nonumber\\ 
& =\langle e^{-p Z(x_0) \Delta t} Q(p, x_0') \rangle.
\label{def-Q1}
\end{align}
To evaluate $x_0'$, note that starting from the initial position $x_0$, in small time $\Delta t$, the particle can either move to $x_0 + \eta (0) \Delta t$ with probability $(1-r \Delta t)$ or reset to $x_R$ with probability $r \Delta t$. Taking contributions from both, we have
\begin{align}
x_0' &=x_0+\eta (0) \Delta t, ~~~~~~~~~~\text{with probability }(1-r \Delta t) , \nonumber \\
& = x_R, ~~~~~~~~~~~~~~~~~~~~~~\text{with probability }r \Delta t
\label{update-resetting}
\end{align}
We now use Eq. \eqref{update-resetting} in Eq. \eqref{def-Q1} along with the noise properties $\langle \eta  (0) \rangle =0$ and $\langle \eta ^2 (0) \rangle =1/\Delta t$ as$\Delta t \to 0$. Keeping only the leading order terms in $\Delta t$ we get the following backward equation in the presence of resetting
\begin{align}
\frac{1}{2} \frac{\partial ^2 Q(p,x_0)}{\partial x_0^2} - p  Z(x_0) Q(p,x_0)-r Q(p,x_0)+r Q(p,x_R) = 0. 
\label{bfp}
\end{align}
Eq. \eqref{bfp} is the central equation of the paper.
Notice simply that it boils down to the result (see Eq. (51)) in \cite{Majumdar2005} for $r=0$. The above equation is valid in the domain $[0,\infty]$ and we need to supplement appropriate boundary conditions to solve it
\begin{align}
& Q(p, x_0 \to 0^+) =1, ~\label{An-eq-2} \\
& Q(p, x_0 \to \infty) < \infty.
\label{An-eq-3}
\end{align}
The first boundary condition is easy to understand since the particle will be absorbed momentarily if it starts from the origin. This means that the first-passage time $t_f \to 0$ (hence the integral $V =\int_0^{t_f}~d\tau~Z[x(\tau)] \to 0$) and from Eq. \eqref{def-Q} one derives Eq. \eqref{An-eq-2}. Let us now look at the second boundary condition. Although the particle is initially at $x_0 \to \infty$, the next resetting event can bring it to the finite location $x_R$ within a finite time window. Therefore, the first-passage time $t_f$ is typically finite for any non-zero and finite value of $r$. Based on this intuition, we expect the functional $V$ to remain finite. This results in the boundary condition in Eq. \eqref{An-eq-3}.

So, given a functional $Z(x)$, the scheme would be to first solve the backward equation Eq. \eqref{bfp} with the
appropriate boundary conditions mentioned above to obtain $Q(p,x_0)$ explicitly and then invert the Laplace transform
in Eq. \eqref{def-Q} to get the desired PDF $P_R(V,x_0)$ of the first-passage functional. In the following sections, we discuss various choices of $Z(x)$ as mentioned in Sec. \ref{introduction} and illustrate the consequences of resetting. 

\begin{figure}[t]
\includegraphics[width=\textwidth]{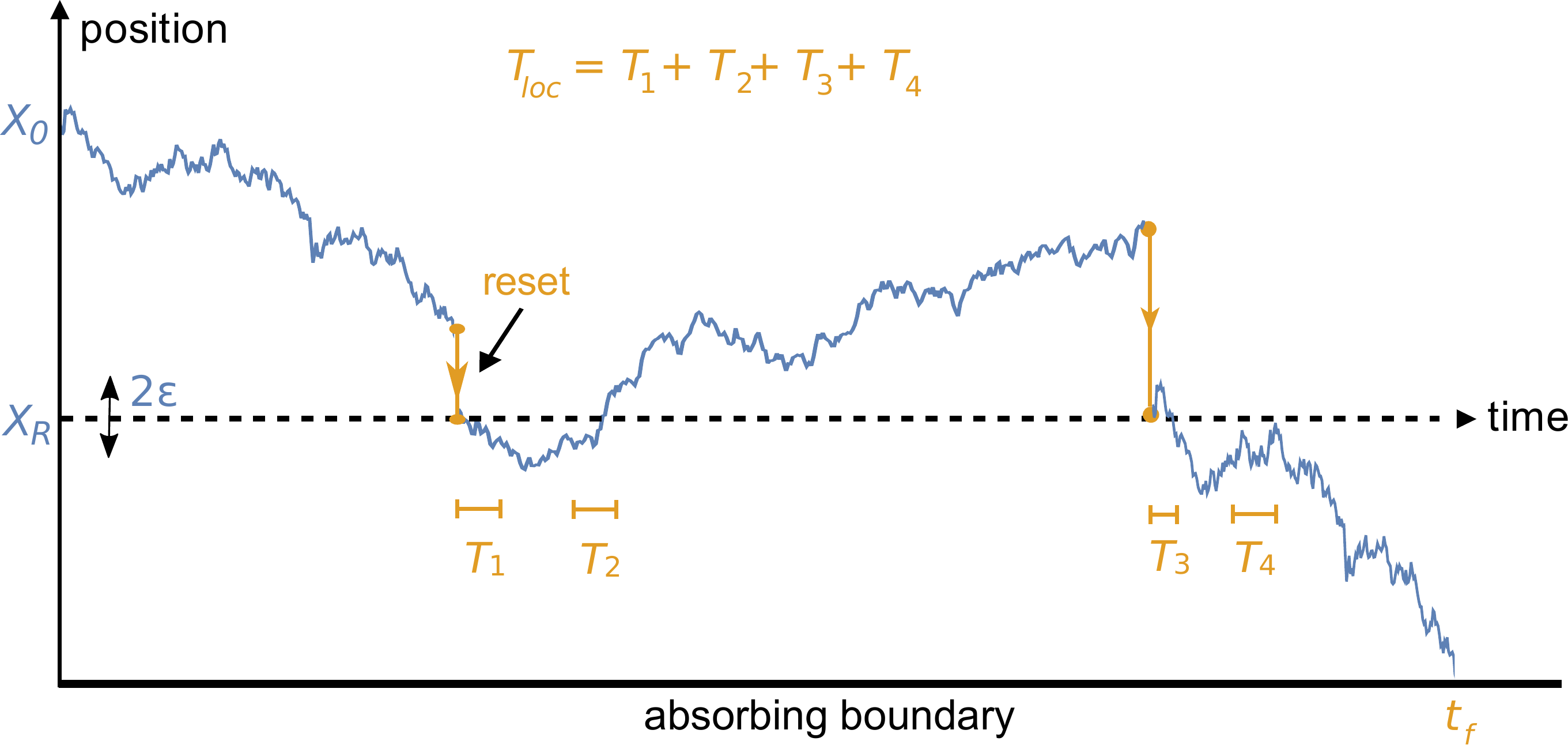}
\centering
\caption{Schematic of local time of diffusion with stochastic resetting. Local time $T_{loc}$ is a collection of all time segments in the trajectory (as shown in the plot) spent by the particle in the domain $2\epsilon$ around $x_R$ upto the first passage time $t_f$.} 
\label{local-scheme}
\end{figure} 

\section{Local time}
\label{local}
The local time density spent 
by the particle at position $x_\ell$ upto the first-passage time $t_f$ is given by
\begin{alignat}{1}
T_{loc}(x_\ell)=\int_{0}^{t_f}d\tau ~\delta[x(\tau)-x_\ell]~.
\label{local-reset}
\end{alignat}
The appearance of the delta function in the above definition is understood in the following limiting sense:
\begin{align}
T_{loc}(x_\ell)= \lim_{\epsilon \to 0}\frac{T_{2\epsilon}(x_\ell)}{2\epsilon},~~~\text{where}~~T_{2\epsilon}(x_\ell)=\int_0^{t_f} d \tau ~[\theta(x(\tau)-x_\ell-\epsilon)-\theta(x(\tau)-x_{\ell}+\epsilon)].
\end{align}
In what follows, such regularization procedure should precede every use of the delta function, but we will omit it for brevity. Clearly, $T_{2\epsilon}(x_\ell)$ measures the total time spent by the particle inside the box $[x_\ell-\epsilon,x_\ell+\epsilon]$ till time $t_f$ in the presence of resetting. Thus, by definition, the normalization condition reads $\int_0 ^{\infty} T_{loc}(x_\ell)dx_\ell=t_f$. Here, we are interested in estimating the local time density near the resetting location $x_R$ [see Fig. \ref{local-scheme}]. Thus, using $Z(x)=\delta (x-x_R)$ from
Eq. \eqref{local-reset} and substituting into the backward Eq. \eqref{bfp}, we obtain
\begin{align}
\frac{1}{2} \frac{\partial ^2 Q(p,x_0)}{\partial x_0^2} - p ~ \delta(x_0-x_R) Q(p,x_0)-r Q(p,x_0)+r Q(p,x_R) = 0. 
\label{BM-loc-eq-1}
\end{align}
For $x_0 \neq x_R$, we get rid of the term with $\delta(x_0-x_R)$ and rewrite Eq. \eqref{BM-loc-eq-1} as
 \begin{align}
\frac{1}{2} \frac{\partial ^2 Q(p,x_0)}{\partial x_0^2} -r Q(p,x_0)+r Q(p,x_R) = 0. 
\label{BM-loc-eq-2}
\end{align}
It is easy to solve this equation and get the solutions
\begin{align}
Q(p,x_0) = \begin{cases}
&\mathcal{A} e^{\sqrt{2r}x_0}+\mathcal{B} e^{-\sqrt{2r}x_0} +Q(p,x_R),~~~~\text{for }x_0 <x_R, \\
&\mathcal{C} e^{\sqrt{2r}x_0}+\mathcal{D} e^{-\sqrt{2r}x_0} +Q(p,x_R),~~~~\text{for }x_0 >x_R.
\end{cases}
\label{BM-loc-eq-3}
\end{align}
Here, the functions $\mathcal{A},~\mathcal{B},~\mathcal{C}$ and $\mathcal{D}$ are independent of $x_0$ but may depend on $x_R$ and $p$. To evaluate these functions, we need to specify four conditions on $Q(p,x_0)$. Two of these conditions come from the behaviour of $Q(p,x_0)$ as $x_0 \to 0^+$ and $x_0 \to \infty$ which are, respectively, written in Eqs. \eqref{An-eq-2} and \eqref{An-eq-3}. In addition, we use two matching
conditions. To see these, we first integrate Eq. \eqref{BM-loc-eq-1} from $x_0 = x_{R}-\delta$ to $x_0 = x_{R}+\delta$ and take $\delta \to 0^+$. Next, we use the continuity of $Q(p,x_0)$ across $x_0 = x_R$. The resultant conditions are
\begin{align}
&~~Q(p,x_R  ^+) = Q(p, x_R  ^-), \label{BM-loc-eq-5}\\
& \left(\frac{\partial Q}{\partial x_0}\right)_{ x_R^+}-\left(\frac{\partial Q}{\partial x_0}\right)_{ x_R^-} = 2 p Q(p, x_R ).\label{BM-loc-eq-6}
\end{align}

\begin{figure}[t]
\includegraphics[scale=0.35]{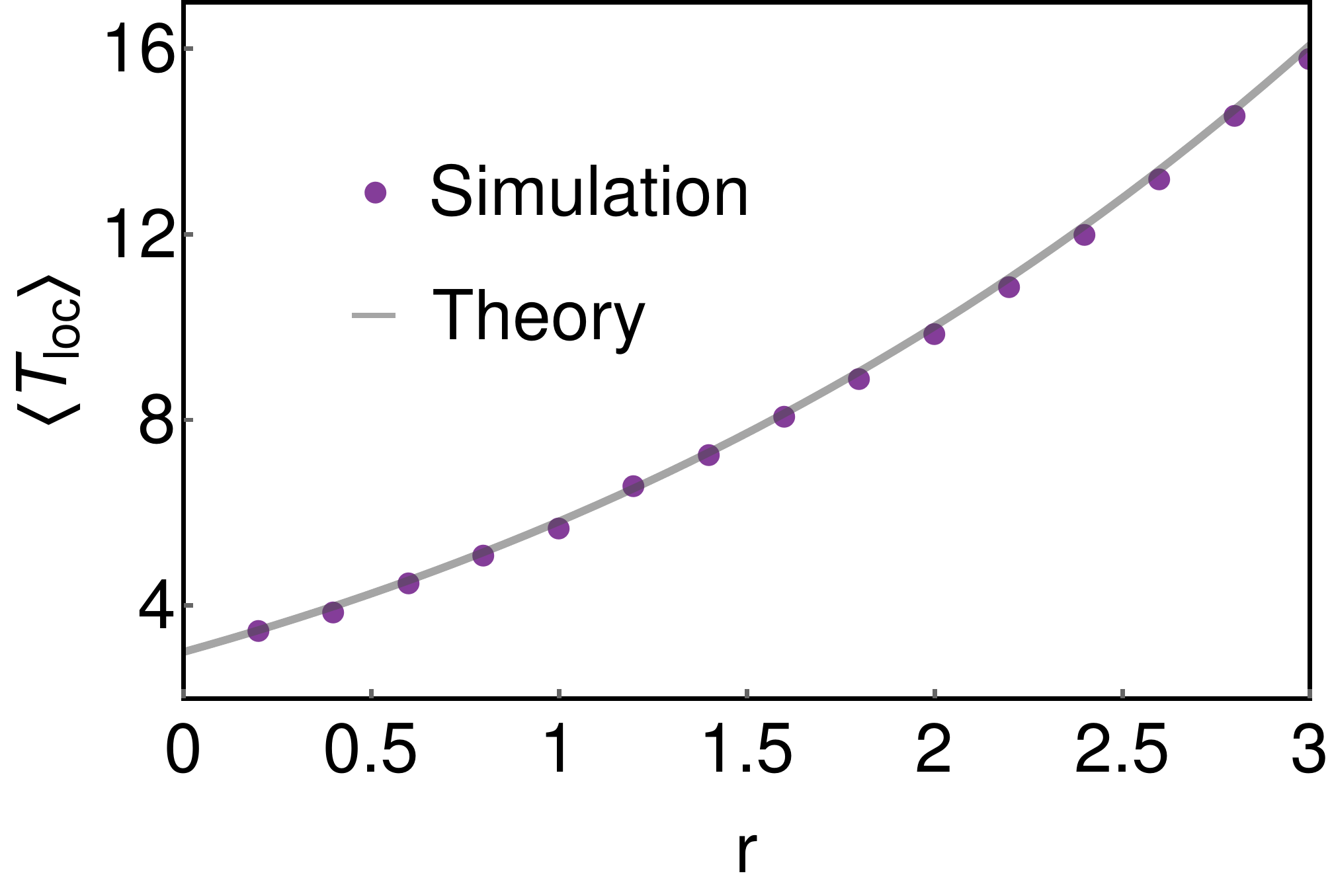}
\includegraphics[scale=0.36]{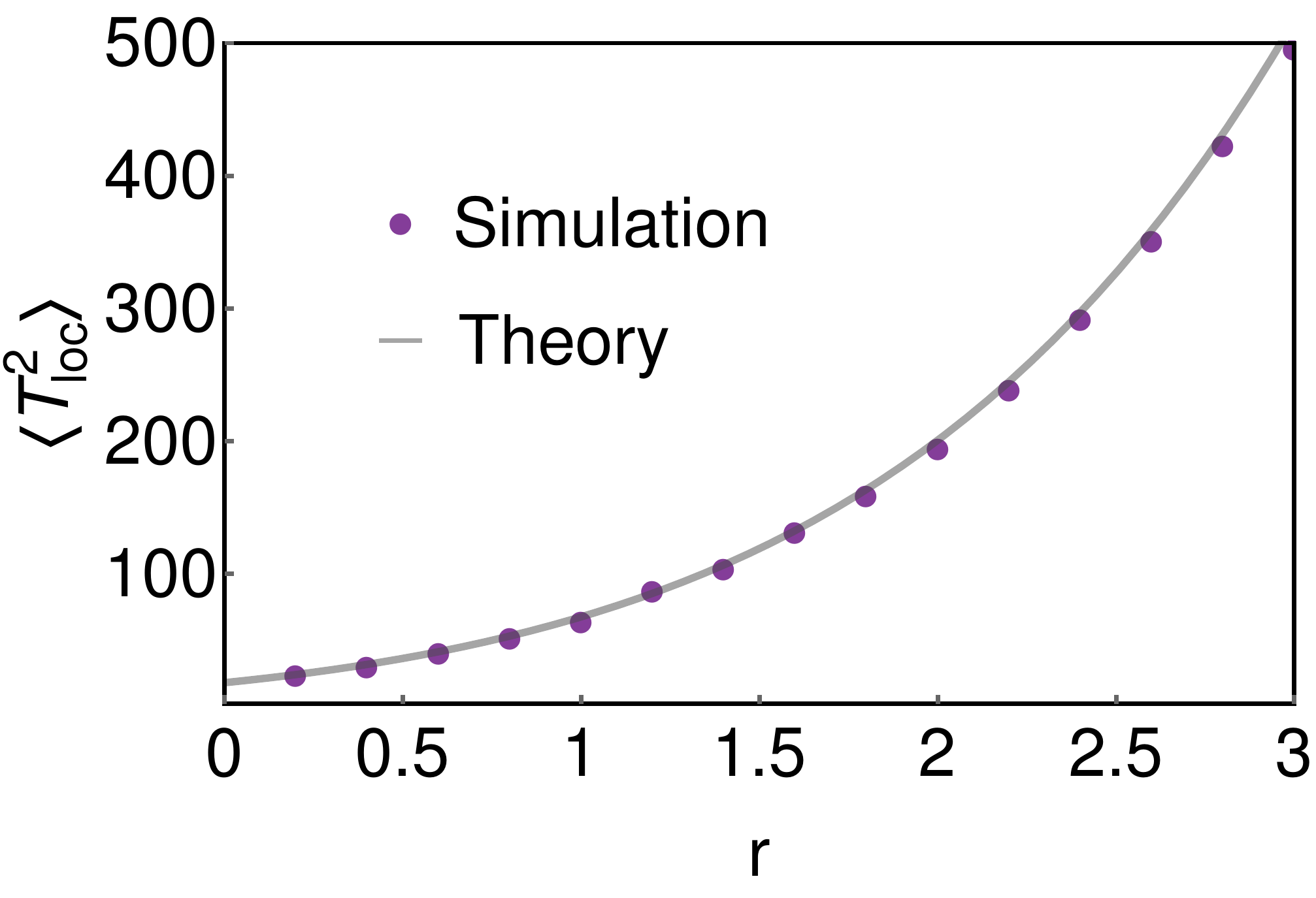}
\centering
\caption{Comparison of the first two moments of $T_{loc}$ for Brownian motion in Eq. \eqref{BM-loc-eq-9} with numerical simulation. We have set $x_0=x_R=1.5$ for both the plots.}  \label{local-moms-Fig}
\end{figure} 

We can now evaluate the functions $\mathcal{A},~\mathcal{B}$ and $\mathcal{D}$ and plug them in Eq. \eqref{BM-loc-eq-3} to obtain the exact form of $Q(p,x_0)$. However, since we are interested in the case $x_0 =x_R$, it is easy to see from Eq. \eqref{BM-loc-eq-3} that $Q(p,x_R)$ is completely characterised by $\mathcal{D}(x_R,p)$ which reads
\begin{align}
\mathcal{D}(x_R,p) = \frac{\sqrt{r} \left[ Q(p,x_R)-1\right]-\sqrt{2} p Q(p,x_R)~ \text{sinh} \left(\sqrt{2r} x_R\right)}{\sqrt{2}\left[ p~\text{sinh} \left(\sqrt{2r} x_R\right) -\sqrt{r}\right]}~e^{\sqrt{2r} x_R}.
\label{BM-loc-eq-7}
\end{align}
Substituting this into the second line of Eq. \eqref{BM-loc-eq-3} yields
\begin{align}
Q(p,x_R) = \frac{\sqrt{r}}{\sqrt{r}+\sqrt{2} p ~\text{sinh} \left(\sqrt{2r} x_R\right)}.
\label{BM-loc-eq-8}
\end{align}
To get the form of the distribution of $T_{loc}$, one has to perform the inverse Laplace transformation of Eq. \eqref{BM-loc-eq-8}. Before this, let us look at the moments of $T_{loc}$ using Eq. \eqref{moms-Q}. Inserting $Q(p,x_R)$ from Eq. \eqref{BM-loc-eq-8} in Eq. \eqref{moms-Q}, we find that the $m$-th order moment of $T_{loc}$ is given by
\begin{align}
\langle T_{loc}^m \rangle = m! \left[\sqrt{\frac{2}{r}} ~\text{sinh} \left(\sqrt{2r} x_R\right) \right]^m.
\label{BM-loc-eq-9}
\end{align}
In Fig. \ref{local-moms-Fig}, we have plotted the first two moments of $T_{loc}$ and compared them against the simulation. We observe an excellent agreement of our analytical results with the simulation. From Eq. \eqref{BM-loc-eq-9}, we find that the local time scales as $T_{loc} \sim \frac{1}{\sqrt{r}}~\text{exp} (\sqrt{2r}x_R)$ for large $r$ which is different than $T_{loc} \sim x_R$ scaling without resetting. As $r$ increases, the particle is brought to $x_R$ more frequently which causes it to spend more time in the vicinity of $x_R$. This results in the enhancement of the local time for non-zero $r$.

\begin{figure}[t]
\includegraphics[scale=0.32]{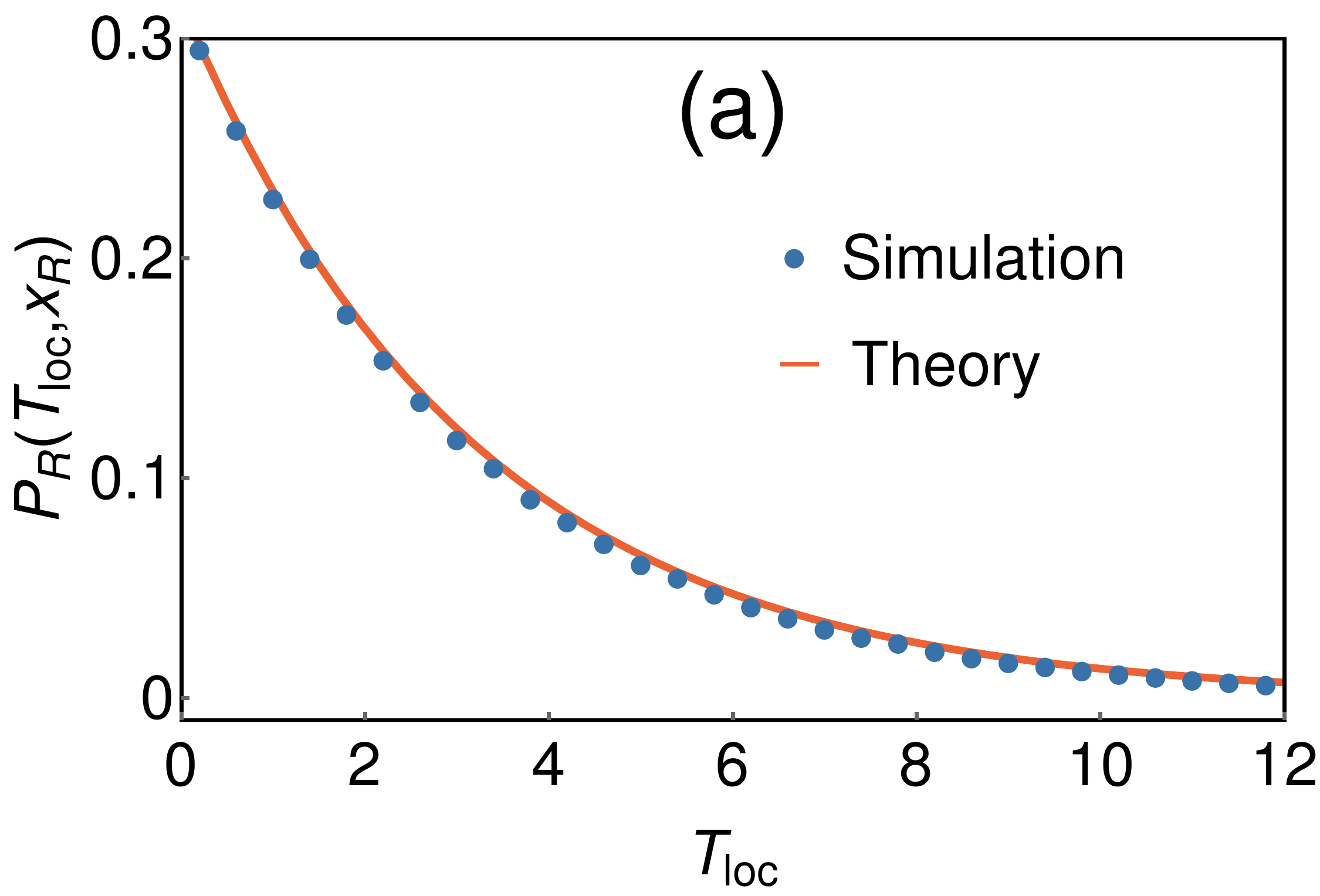}
\includegraphics[scale=0.31]{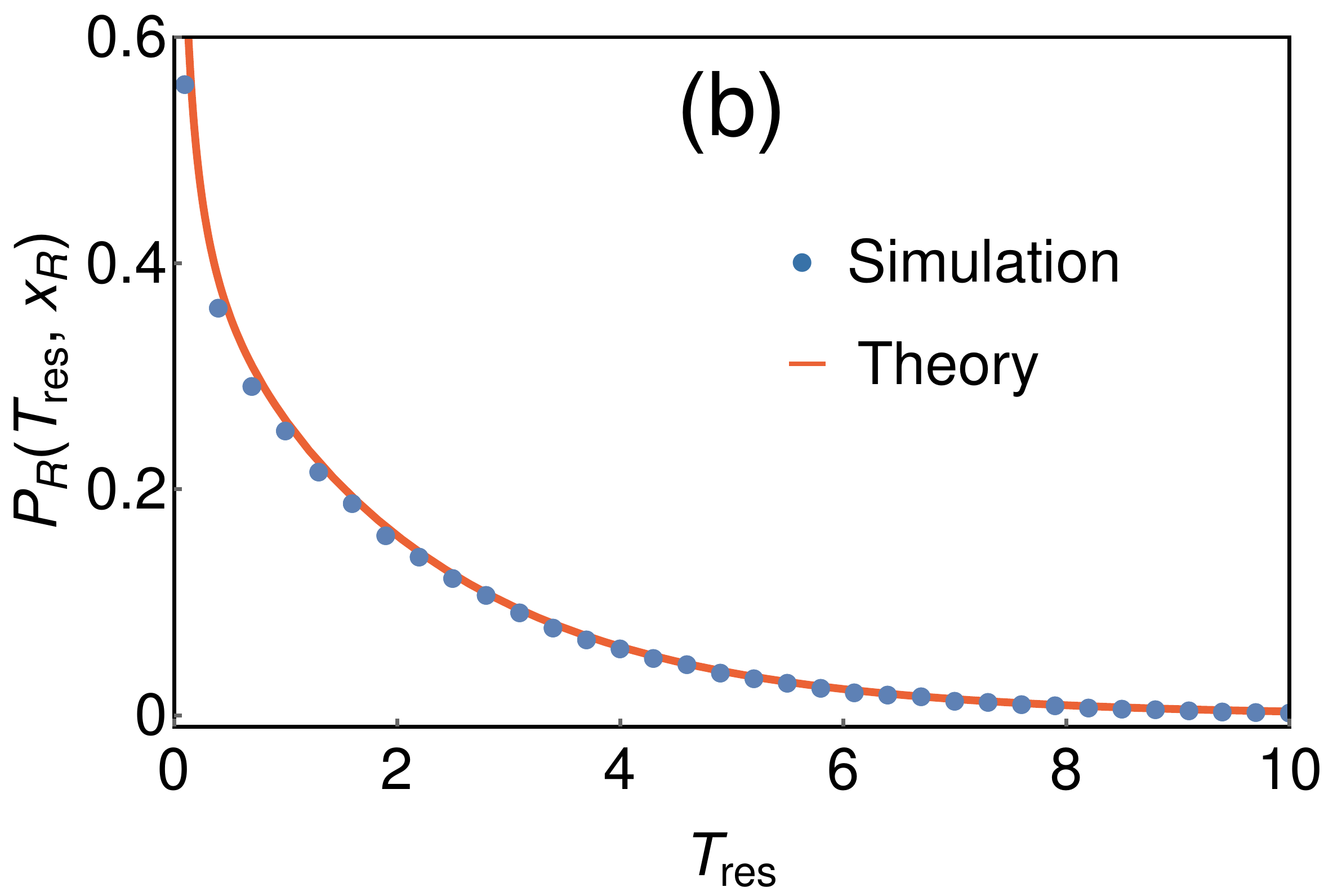}
\centering
\caption{(a) Theoretical plot of $P_R(T_{loc}, x_R)$  (Eq. \eqref{BM-loc-eq-10}) and its comparison with the numerical simulations. Parameters chosen are $x_0=x_R=1$ and $r=1.5$. (b) Comparison of the residence time distribution $P_R(T_{res}, x_R)$ in Eq. \eqref{BM-occ-eq-13} with the numerical simulation. Here, we have set $x_0=x_R=1$ and $r=2$.} 
\label{local-dist-Fig}
\end{figure} 

Returning to the distribution of $T_{loc}$, we next perform the inverse Laplace transformation of $Q(p,x_R)$ in Eq. \eqref{BM-loc-eq-8} and obtain
\begin{align}
P_R\left(T_{loc},x_R \right) = \frac{\sqrt{r}}{{\sqrt{2} ~\text{sinh} \left(\sqrt{2r} x_R\right)}}~\text{exp} \left( -\frac{\sqrt{r}}{{\sqrt{2} ~\text{sinh} \left(\sqrt{2r} x_R\right)}}T_{loc}\right).
\label{BM-loc-eq-10}
\end{align}
In Fig. \ref{local-dist-Fig}(a), we have plotted $P_R\left(T_{loc},x_R \right)$ and compared with the results of numerical simulations. The distribution $P_R\left(T_{loc},x_R \right) $ decays exponentially with decay length equal to $\sqrt{2}\sinh\left(\sqrt{2r} x_R\right) /\sqrt{r}$. Since the typical value of $T_{loc}$ increases with $r$, the decay length also increases with $r$.

\section{Residence time}
\label{residence}
\begin{figure}[t]
\includegraphics[width=\textwidth]{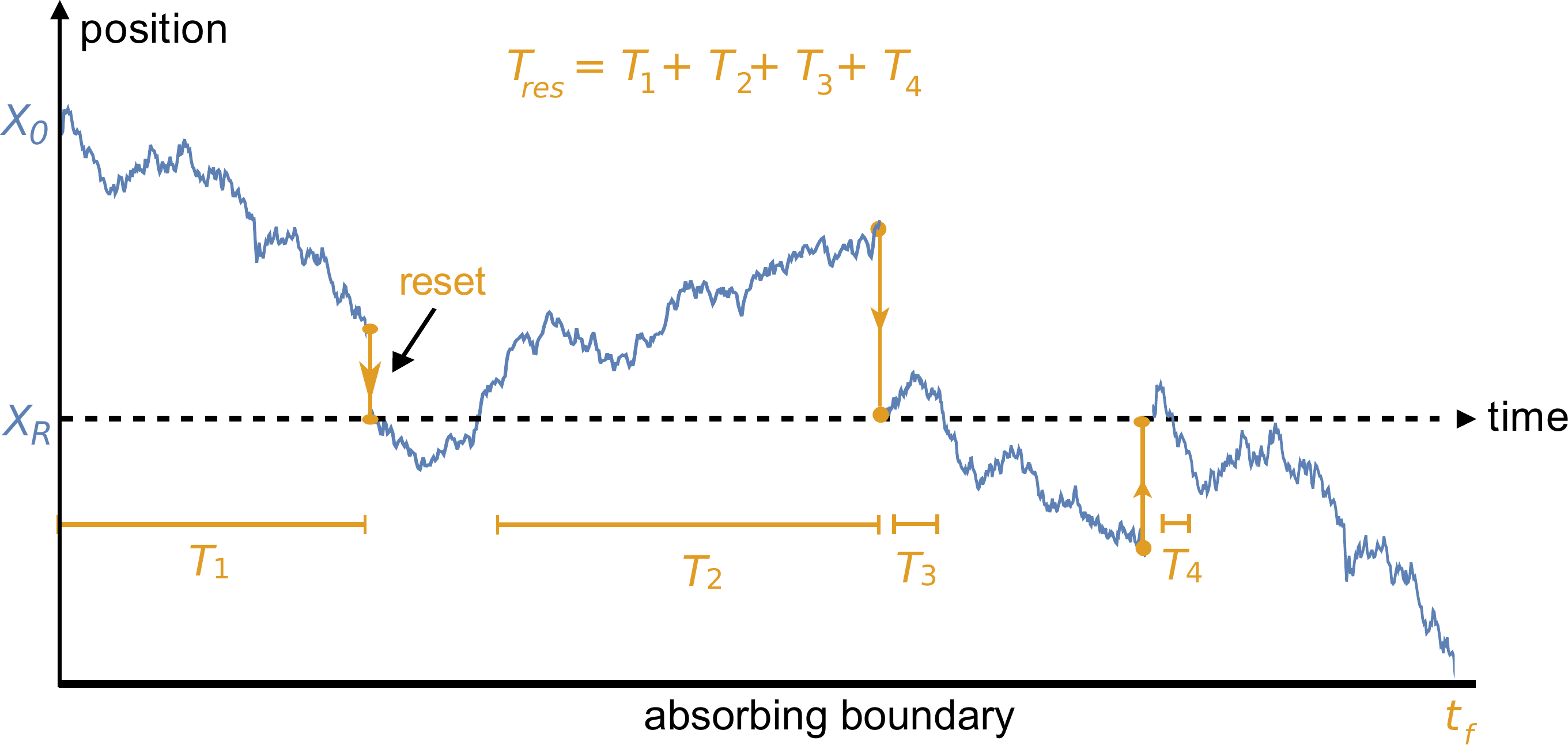}
\centering
\caption{Schematic of residence time of diffusion with stochastic resetting. Residence time $T_{res}$ is a collection of all time segments in the trajectory (as shown in the plot) spent by the particle in the domain $x>x_R$ upto the first passage time $t_f$.} 
\label{residence-scheme}
\end{figure}

In this section, we discuss 
the effects of resetting on the statistics of residence/occupation time $T_{res}$. We define residence time as the cumulative time that particle spends in the region $x > x_R$ before getting absorbed at the origin, thus $T_{res}=\int_0^{t_f}~d\tau~\theta(x(\tau)-x_R)$ [see Fig. \ref{residence-scheme}]. Substituting $Z(x) = \theta(x-x_R)$ into Eq. \eqref{bfp}, we get
\begin{align}
\frac{1}{2} \frac{\partial ^2 Q(p,x_0)}{\partial x_0^2} - p  \theta (x_0-x_R) Q(p,x_0)-r Q(p,x_0)+r Q(p,x_R) = 0. 
\label{BM-occ-eq-1}
\end{align}
Solving this equation separately for $x_0 \geq x_R$ (for which $\theta (x_0-x_R)=1$) and  $x_0 < x_R$ (for which $\theta (x_0-x_R)=0$), we get
\begin{align}
Q(p,x_0) = \begin{cases}
&\mathcal{C}_1 e^{\sqrt{2r}x_0}+\mathcal{C}_2 e^{-\sqrt{2r}x_0} +Q(p,x_R),~~~~~~~~~\text{for }x_0 <x_R,\\
&\mathcal{C}_3 e^{\sqrt{2r}x_0}+\mathcal{C}_4 e^{-\sqrt{2r}x_0} +\frac{r}{r+p}Q(p,x_R),~~~~~\text{for }x_0 >x_R .
\end{cases}
\label{BM-occ-eq-2}
\end{align}
The constants $\mathcal{C}_1,~\mathcal{C}_2,~\mathcal{C}_3$ and $\mathcal{C}_4$ are independent of $x_0$ but are, in general, functions of $p$ and $x_R$. To compute them, we use the boundary conditions in Eqs. \eqref{An-eq-2} and \eqref{An-eq-3} along with the continuity conditions for $Q(p,x_0)$ and $\frac{\partial Q(p,x_0)}{\partial x_0}$ across $x_0 = x_R$
\begin{align}
& Q(p, x_R^+) = Q(p, x_R^-),\label{BM-occ-eq-5}\\
& \left( \frac{\partial Q}{\partial x_0} \right)_{x_R^+}=\left( \frac{\partial Q}{\partial x_0} \right)_{x_R^-}.\label{BM-occ-eq-6}
\end{align}
Using these conditions, it is straightforward to compute all $\mathcal{C}(p,x_R)$ functions in Eq. \eqref{BM-occ-eq-2}. However, once again, we are interested in the case where $x_0=x_R$. For this, we need to specify only the expression of  $\mathcal{C}_4(p,x_R)$ as indicated by the second line in Eq. \eqref{BM-occ-eq-2} [$\mathcal{C}_3(p,x_R)=0$ from Eq. \eqref{An-eq-3}]. The expression of $\mathcal{C}_4(p,x_R)$ reads
\begin{align}
\mathcal{C}_4(p,x_R) = \frac{\left[1-Q(p,x_R) \right] (p+r)+p~ Q(p,x_R) \text{cosh} \left( \sqrt{2r} x_R\right)}{(p+r)\left[ \text{cosh} \left( \sqrt{2r} x_R\right)-\sqrt{\frac{p+r}{r}}~\text{sinh} \left( \sqrt{2r} x_R\right)\right]}~e^{\sqrt{2(r+p)} x_R}.
\label{BM-occ-eq-7}
\end{align}
and inserting this in the second line of Eq. \eqref{BM-occ-eq-2} gives
\begin{align}
Q(p, x_R) = \frac{\sqrt{r(r+p)}}{\sqrt{r(r+p)} + p ~\text{sinh} \left( \sqrt{2r} x_R\right)}.
\label{BM-occ-eq-8}
\end{align}

\begin{figure}[t]
\includegraphics[scale=0.35]{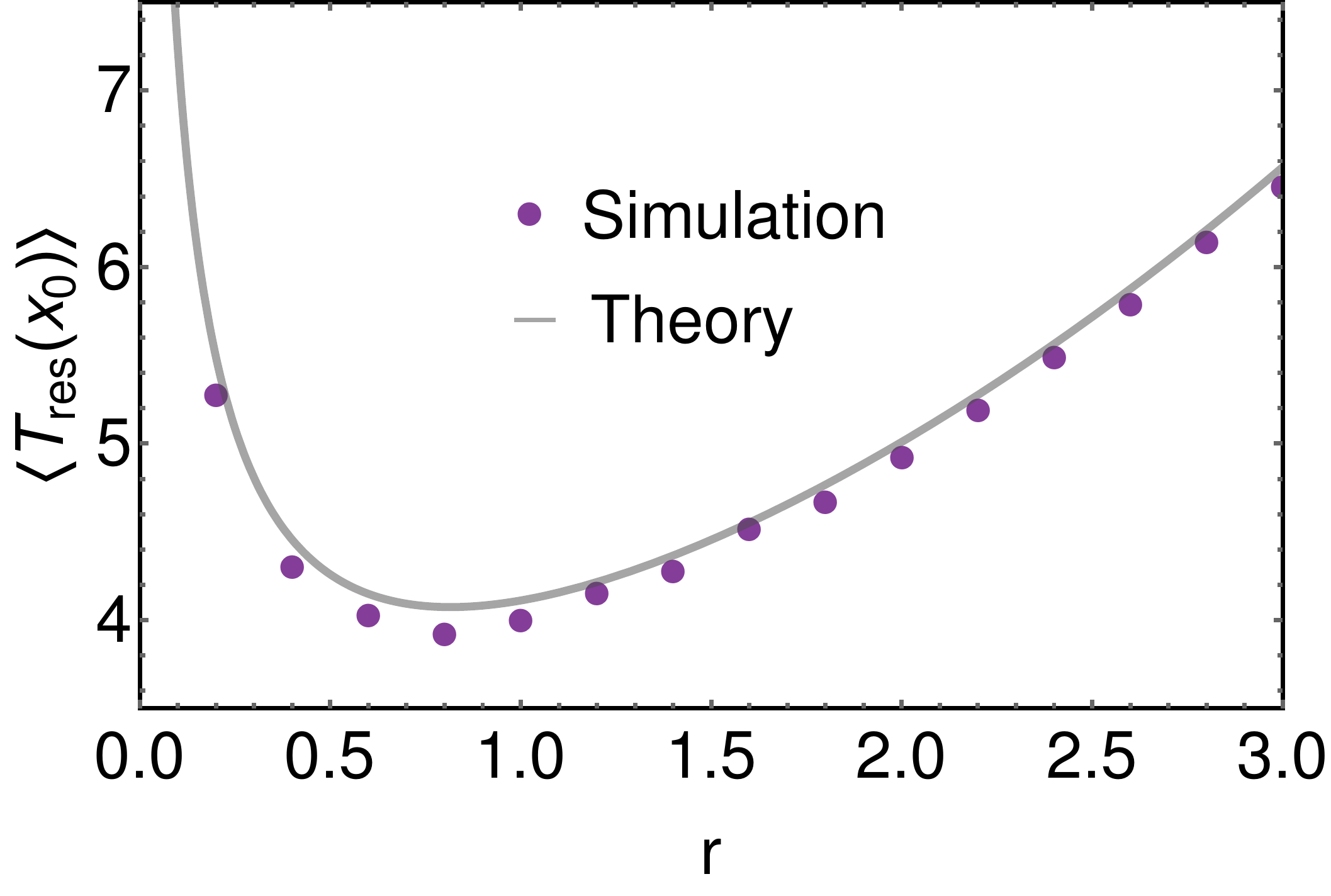}
\includegraphics[scale=0.35]{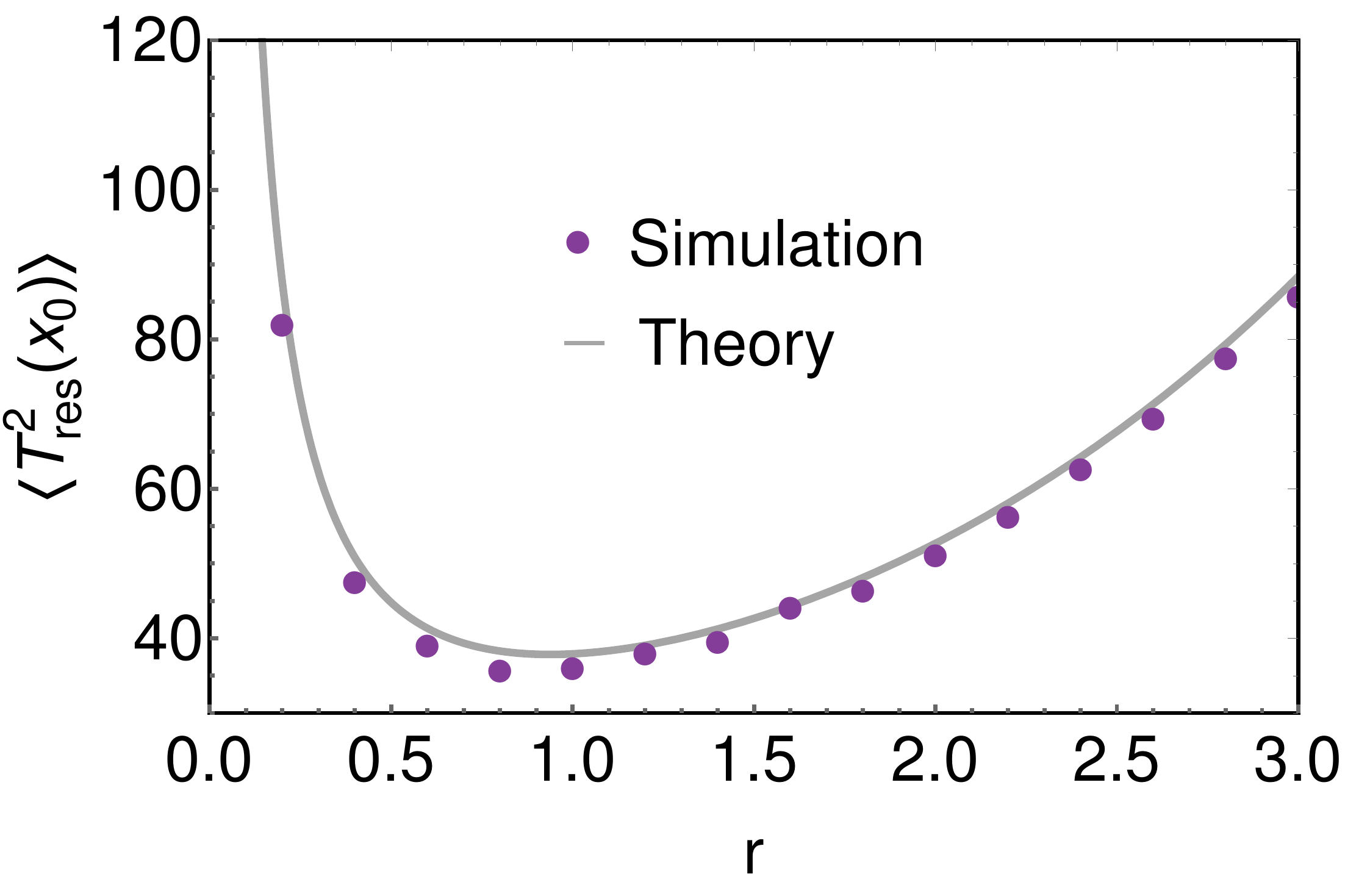}
\centering
\caption{Comparison of the first two moments of $T_{res}$ given by Eqs. \eqref{BM-occ-eq-9} and \eqref{BM-occ-eq-10} with numerical simulations. Here, we have set $x_0=x_R=1.5$ in both the plots.}  
\label{occ-moms-Fig}
\end{figure}

To summarise, we have obtained the exact form of the Laplace transformation of the distribution $P_R(T_{res},x_R)$ from which the moments can be computed using Eq. \eqref{moms-Q}. 
Here, we provide exact expressions for the first two moments:
\begin{align}
&\langle T_{res}(x_R) \rangle = \frac{\sinh \left( \sqrt{2r} x_R\right)}{r}, \label{BM-occ-eq-9} \\
&\langle T_{res} ^2(x_R)\rangle = \frac{\sinh\left( \sqrt{2r} x_R\right) }{r^2}\left[ 1+2 \sinh \left( \sqrt{2r} x_R\right) \right]. \label{BM-occ-eq-10} 
\end{align}
These two moments are plotted in Fig. \ref{occ-moms-Fig} where we have also performed comparison with the numerical simulations. In the limit $r\to 0$, the moments of $T_{res}$ diverge. On the other hand, repeated resetting at a rate $r$ renders the moments finite. Physically, this can be understood in the following way: For standard Brownian motion, there will be some trajectories for which the particle will always stay on $x > x_R$ region and avoid getting absorbed at the origin. Such trajectories contribute to $t_f \to \infty$ and $T_{res} \to \infty$. From an analysis shown later, we find that the weight of these trajectories to the distribution is $\sim 1/ \sqrt{T_{res}}$. This results in the diverging moments. However, in presence of resetting, particle will be brought to $x_R$ intermittently which will reduce drastically the effective contribution of trajectories with $t_f \to \infty$ and $T_{res} \to \infty$. Under this situation, the effective contribution of these trajectories to the probability distribution is $\sim e^{-a T_{res}}$ (see  below) where $a$ is some positive function of $r$ and $x_R$. This, in turn, results in finite moments.

Quite interestingly, we observe that the moments exhibit a non-monotonic behaviour with respect to $r$ with minimum value at some optimal resetting rate $r^*$(see Fig. \ref{occ-moms-Fig}). We saw above that the first two moments diverge respectively as $\sim 1/ \sqrt{r}$ and $\sim 1/ r^{3/2}$ for $r \ll r^*$. Hence, the moments decrease with increasing $r$ for $r \ll r^*$. On the other hand, for $r \gg r^*$, the particle is frequently reset to $x_R$ which reduces its likelihood for a first passage to the origin. As a result, one expects the typical value of residence time $T_{res}$ to increase with $r$ for $r \gg r^*$. In fact, Eqs. \eqref{BM-occ-eq-9} and \eqref{BM-occ-eq-10} reveal that the first two moments diverge as $\sim r^{-1} e^{\sqrt{2 r} x_R}$ and $\sim r^{-2} e^{2\sqrt{2 r} x_R}$ respectively for $r \to \infty$. Therefore, at some $r=r^*$, the slope of the moments changes from negative to positive which essentially corresponds to their minimum value.

After analysing the moments, we next proceed to compute the distribution of $T_{res}$. To this end, we rewrite $Q(p, x_R)$ as
\begin{align}
Q(p, x_R) = \sum _{k=0}^{\infty} \frac{(-1)^k ~r^{\frac{k+1}{2}}}{\left(  \sinh \left( \sqrt{2r} x_R\right)\right)^{k+1}}~\left(\frac{\sqrt{p+r}}{p} \right)^{k+1}.
\label{BM-occ-eq-11} 
\end{align}  
To proceed further, we use the following inverse Laplace transformation:
\begin{align}
\mathcal{L}^{-1}_{p \to T_{res}} \left[ \left(\frac{\sqrt{p+r}}{p} \right)^{k+1}.\right]=T_{res}^{\frac{k-1}{2}} ~_1\bar{F}_1 \left( -\frac{k+1}{2},\frac{k+1}{2}, -r T_{res}\right),~~~\text{for }k>-1,
\label{BM-occ-eq-12} 
\end{align}
where $\mathcal{L}^{-1}_{p \to T_{res}} \left[ g(p)\right]$ denotes an inverse Laplace transformation of the function $g(p)$. Here, $_1\bar{F}_1(a,b,y)$ stands for the regularized confluent hypergeometric function \cite{eqs}. Finally, from Eq. \eqref{BM-occ-eq-11}, we have 
\begin{align}
P_R(T_{res},x_R) = \frac{r}{\sinh \left( \sqrt{2r} x_R\right)} ~H(r T_{res}),
\label{BM-occ-eq-13} 
\end{align}
where the scaling function $H(y)$ is given by
\begin{align}
H(y) = \frac{1}{\sqrt{y}}\sum _{k=0}^{\infty} \left(-\frac{\sqrt{y}}{\sinh \left( \sqrt{2r} x_R\right)} \right)^{k}~_1\bar{F}_1 \left( -\frac{k+1}{2},\frac{k+1}{2}, -y\right).
\label{BM-occ-eq-14}
\end{align}
In Fig. \ref{local-dist-Fig}(b), we have plotted $P_R(T_{res},x_R)$ and compared it with the numerical simulations to find an excellent match. To see the effect of resetting on the distribution, it is instructive to look at the asymptotic form of $H(y)$. As $y \to 0$, we have $~_1\bar{F}_1 \left( -\frac{k+1}{2},\frac{k+1}{2}, -y \right) \simeq \frac{1}{\Gamma\left( \frac{k+1}{2}\right)}$. On the other hand for large $y$, we have $~_1\bar{F}_1 \left( -\frac{k+1}{2},\frac{k+1}{2}, -y \right) \simeq \frac{(\sqrt{y})^{k+1}}{\Gamma\left( k+1\right)}$. Inserting these asymptotic forms in Eq. \eqref{BM-occ-eq-14} and performing some algebraic simplifications, we find
\begin{align}
H(y) &\simeq \frac{1}{\sqrt{\pi y}},~~~~~~~~~~~~~~~~~~~\text{as }y \to 0,\label{BM-occ-eq-15}\\
& \simeq e^{-\frac{y}{\sinh \left( \sqrt{2r} x_R\right)}},~~~~~~~~~~~\text{as }y \to \infty.
\label{BM-occ-eq-16}
\end{align}
Finally, substituting these forms in Eq. \eqref{BM-occ-eq-13}, we find that the distribution $P_R(T_{occ}, x_R)$ has the following asymptotic forms:
\begin{align}
P_R(T_{res}, x_R) & \simeq \frac{\sqrt{r}}{\sqrt{\pi T_{res}}~\sinh \left( \sqrt{2r} x_R\right)},~~~~~~~~~~~~~~~~~~~~~~~~~~~~~~\text{for }T_{res} \ll r^{-1}, \label{BM-occ-eq-17}\\
&\simeq \frac{r}{\sinh \left( \sqrt{2r} x_R\right)} \text{exp}\left(-\frac{r ~T_{res}}{\sinh \left( \sqrt{2r} x_R\right)} \right),~~~~~~~~\text{for }T_{res} \gg r^{-1}. \label{BM-occ-eq-18}
\end{align}
We see that for no-resetting case, the distribution exhibits power-law decay of the form $\sim 1/ \sqrt{T_{res}}$. On the other hand, in presence of resetting, the distribution has an exponential tail of the form in Eq. \eqref{BM-occ-eq-18}. Emergence of exponential tails is a hallmark property for observables under stochastic resetting mechanism which was first noted in the context of first passage time \cite{Restart1} (also see below and \cite{local-r}). Here too, we observe a similar behavior which essentially results in finite moments in comparison to the underlying process.

\section{Functional of the form $A_n (x_0)= \int _{0}^{t_f}~[x(\tau)]^n d \tau$}
\label{area}
In this section, we focus on functionals of type 
\begin{align}
A_n(x_0) = \int _{0}^{t_f}~[x(\tau)]^n d \tau,~~~~ \text{with} ~~~n >-2.    
\end{align}
As discussed before, $A_n(x_0)$ for different values of $n$ arises in various contexts like search problem, queueing theory, particle in harmonic or Sinai potential. Statistics of $A_1(x_0)$ (for the $r=0$ case) was studied in different contexts like sandpile model and directed peroclation  \cite{Kearney2005, Kearney2004, Prellberg1995, Majumdar2007, Dhar1989}. For $r=0$, the distribution of $A_n(x_0)$ was exactly computed in \cite{Meerson2020} for all $n >-2$. Here, we study this problem in presence of resetting. In particular, note that for $n=0$, one has $A_0(x_0) = t_f$ which is nothing but the first-passage time of diffusion with resetting. The topic has been studied quite extensively spanning different stochastic processes that undergo resetting \cite{Evansrev2020,Restart7,Restart8,Restart11,Restart-Search2,Restart-Search3,Restart-Search5,Restart-Search6,RAP,RAP-2,Optimization}. However, to the best of our knowledge, the $n \neq 0$ cases have not yet been studied. This section provides a systematic understanding of the problem for general $n$.

As before, the starting point is to substitute the form of the function i.e., $Z(x)=x^n$ into the Eq. \eqref{bfp}
\begin{align}
\frac{1}{2} \frac{\partial ^2 Q(p,x_0)}{\partial x_0^2} - p ~ x_0^n~ Q(p,x_0)-r Q(p,x_0)+r Q(p,x_R) = 0,
\label{An-eq-1}
\end{align}
which is to be solved along with boundary conditions in Eqs. \eqref{An-eq-2} and \eqref{An-eq-3}. Solving this equation for arbitrary $n$ turns out to be difficult. Only for $n=0$ and $n=1$, we can exactly solve Eq. \eqref{An-eq-1}. For other values of $n$, we present some results on the moments and distribution based on heuristic analysis. In what follows, we briefly discuss the case of $n=0$ for the sake of completeness and then look at the $n=1$ and general $n~(>-2)$ cases separately.
\subsection{$n=0$}
\begin{figure}[t]
\includegraphics[scale=0.38]{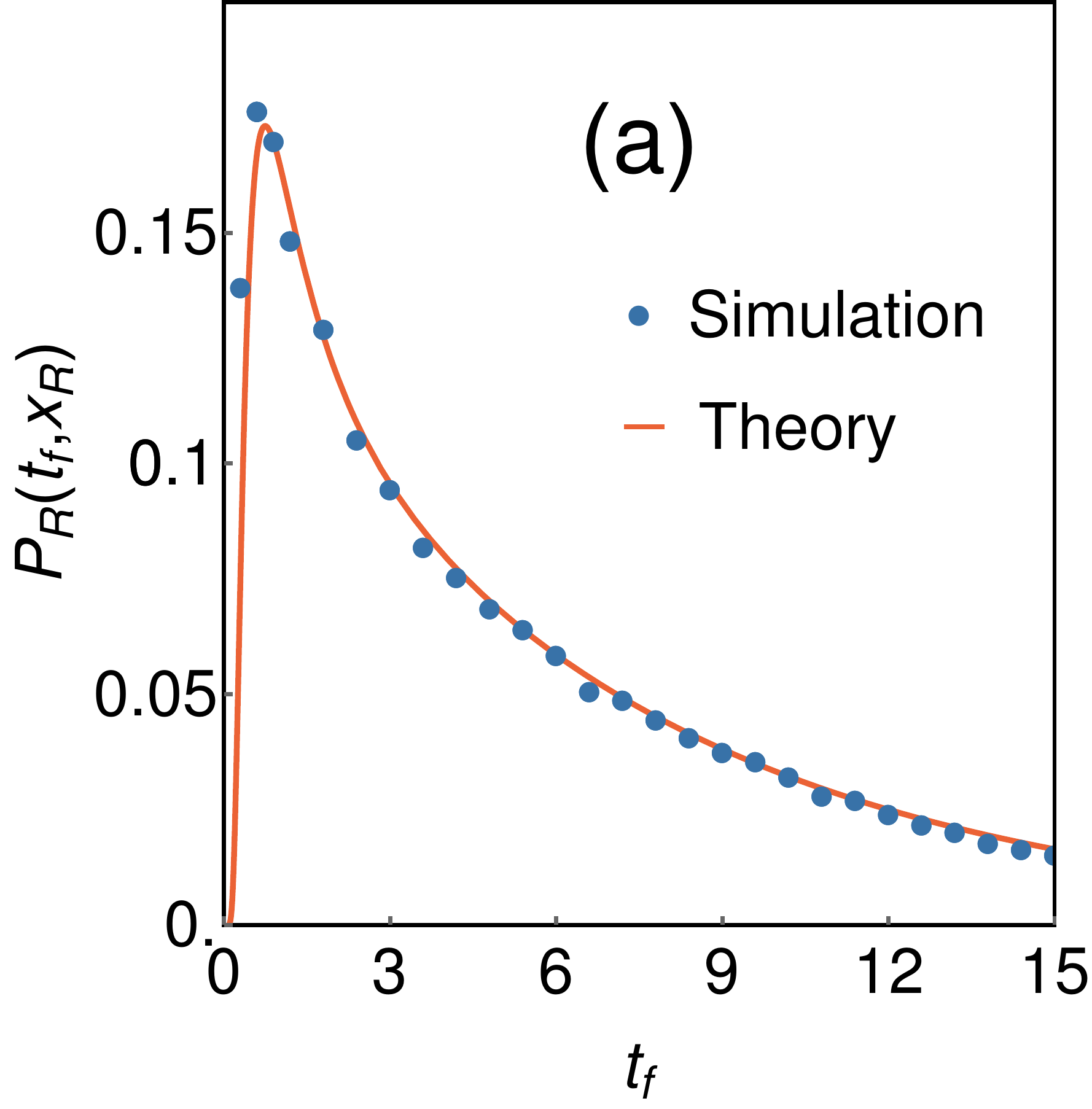}
\includegraphics[scale=0.32]{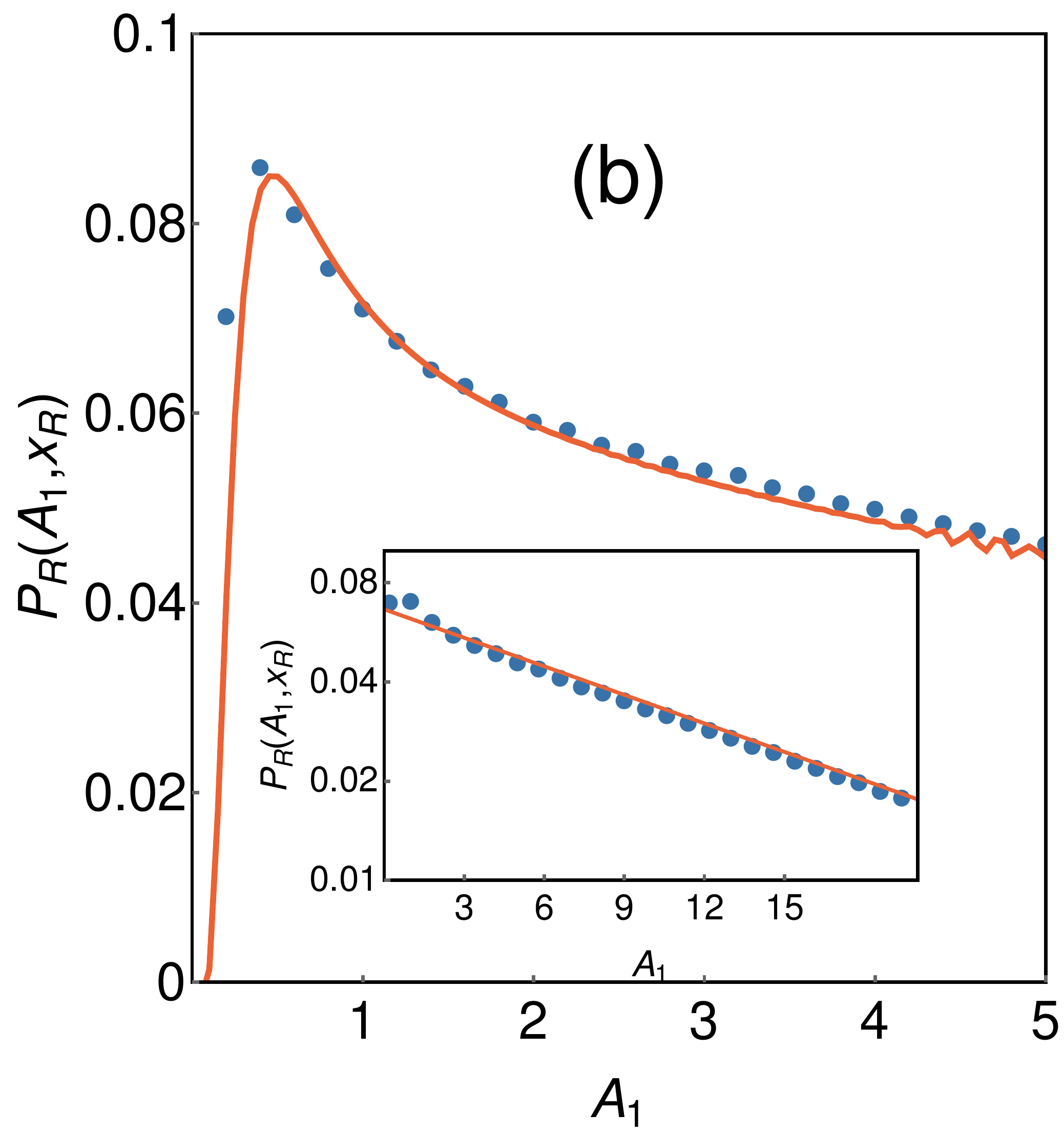}
\centering
\caption{Left panel: analytical result for the distribution $P_R(t_f, x_R)$ in Eq. \eqref{surv-no-drift-w-reset} and its comparison with the simulation. We have set $r=0.5$ and $x_0 =x_R=1.5$. Right panel: plot of the distribution $P_R(A_n, x_R)$ (solid line) for $n=1$ by numerically inverting the Laplace transform $Q(p,x_R)$ in Eq. \eqref{An-eq-13}. We have compared it with the simulation data (circles). \textit{Inset}: plot of the asymptotic form of $P_R(A_1, x_R)$ for large $A_1$ in Eq. \eqref{An-eq-18} (solid line) and its comparison with the simulation results (circles). Parameters taken are $x_0=x_R=1.5$ and $r=2$.}     
\label{area-dist-Fig}
\end{figure}

Let us first look at the case $n=0$ for which $A_0(x_0)$ represents the first passage time (FPT) $t_f$ to the origin starting from $x_0$. It is easy to solve Eq. \eqref{An-eq-1} for this case and obtain
\begin{align}
Q(p,x_0) = \mathbb{C}_1 e^{-\sqrt{2(r+p)} x_0}+\mathbb{C}_2 e^{\sqrt{2(r+p)} x_0}+\frac{r}{r+p} Q(p, x_R),
\label{An-eq-4}
\end{align}
where $\mathbb{C}_1$ and $\mathbb{C}_2$ are constants which may be functions of $p$ and $x_R$. We next use the boundary conditions in Eqs. \eqref{An-eq-2} and \eqref{An-eq-3} to evaluate these constants. Finally taking $x_0=x_R$, we get
\begin{align}
Q(p,x_R) = \frac{(r+p)~e^{-\sqrt{2(r+p)} x_R}}{p+r~e^{-\sqrt{2(r+p)} x_R}}.
\label{An-eq-5}
\end{align}
The mean first passage time can be obtained simply as 
\begin{align}
    \langle t_f \rangle = -\partial_p Q(p,x_R)|_{p \to 0}= \frac{1}{r}\left( e^{\sqrt{2r}x_R}-1 \right)~,
\end{align}
which was first derived by Majumdar and Evans in \cite{Restart1}. The first passage time (under resetting) density can be exactly computed by inverting the Laplace inversion in Eq. \eqref{An-eq-5}. We refer the readers to \ref{ILT} for the complete derivation. The final expression reads 
\begin{align}
P_R(t_f,x_R) & = \left[\frac{r u_0 ~e^{-b_0 \sqrt{u_0}}}{1-\frac{b_0}{2 \sqrt{u_0}} e^{- b_0 \sqrt{u_0}}} \right]e^{-r(1-u_0) t_f}+ H_r(t_f), ~~~~~~~~\text{with}\label{surv-no-drift-w-reset}\\
 H_r(t_f) & = \frac{e^{-r t_f}}{\pi} \int _{0}^{\infty} dw ~e^{-w t_f}~\frac{w(r+w) \sin \left(\sqrt{2 w} x_R \right)}{(r+w)^2+r^2-2 r(r+w) \cos \left(\sqrt{2 w} x_R \right)}
\end{align}
where $u_0 ~(0 \le u_0 \leq 1)$ is a solution of the equation $u_0=1-\exp \left( -\sqrt{u_0}~ b_0\right)$ and $b_0=\sqrt{2r}x_R$. For $r=0$, the first term in Eq.~\eqref{surv-no-drift-w-reset} vanishes. Performing integration over $w$ in $ H_r(t_f)$, we get
\begin{align}
P_R(t_f,x_R) = \frac{1}{\sqrt{2 \pi t_f^{3/2}}} ~\text{exp}\left(-\frac{x_R^2}{2 t_f}\right),~~~\text{for }r = 0 
\end{align}
which reduces to the well known result for Brownian motion \cite{Redner2001}. On the other hand, in the large $t$ limit, the first term in Eq.~\eqref{surv-no-drift-w-reset} dominates and the first-passage time distribution decays as $P_R(t_f,x_R) \sim e^{-r(1- u_0 )t_f}$ [the long time behavior can also be understood from the the extreme value statistics (see \cite{Restart1,EVS}) and the survival probability \cite{local-r}]. This is noteworthy since the tails of the FPT distribution are exponential while the underlying diffusion has power-law ($\sim t^{-3/2}$) first passage tails \cite{Redner2001}. 
Simply put, resetting mechanism mitigates large time fluctuations that could induce delays. In fact, more the fluctuations, the better it works thus turning a marked drawback into a favourable advantage (see \cite{inspection} for a very pedagogic viewpoint on this issue). For completeness, we have compared the analytic expression in Eq. \eqref{surv-no-drift-w-reset} against the numerical simulations in Fig. \ref{area-dist-Fig} (left panel).

\subsection{$n=1$}
Let us now turn to another exactly solvable case with $n=1$. Recall that for this case, $A_1(x_0)$ represents the total area swept by the particle till it gets absorbed. To solve Eq. \eqref{An-eq-1} for $n=1$, we make the following transformations: 
\begin{align}
& \bar{Q}(p, x_0) = \frac{p^{2/3}}{\pi r 2^{1/3}} \left[\frac{Q(p,x_0)}{Q(p, x_R)} \right], \label{An-eq-7}\\
& z =(2p)^{\frac{1}{3}} \left(x_0+\frac{r}{p} \right). \label{An-eq-8}
\end{align}
Rewriting Eq. \eqref{An-eq-1} in terms of these variables
\begin{align}
\frac{\partial^2 \bar{Q}(p, z) }{\partial z^2}-z \bar{Q}(p, z) = -\frac{1}{\pi}.
\label{An-eq-9}
\end{align}
The solution of this equation is given in terms of the Scorer's function $\text{Gi}(z)$ as \cite{NIST,eqs} 
\begin{align}
\bar{Q}(p, z) = \mathbb{C}_3 \text{Ai}(z)+ \mathbb{C}_4 \text{Bi}(z) + \text{Gi}(z),
\end{align}
where $\text{Ai}(y)$ and $\text{Bi}(y)$ are Airy functions. Retracing back to the actual variables $x_0$ and $Q(p, x_0)$ in Eqs. \eqref{An-eq-7} and \eqref{An-eq-8}, we get
\begin{align}
& Q(p, x_0) = \mathbb{C}_3 \text{Ai}\left(y_p(x_0) \right)+\mathbb{C}_4 \text{Bi}\left(y_p(x_0) \right) + \frac{\pi r 2^{1/3}}{p^{2/3}} Q(p, x_R)~\text{Gi}\left(y_p(x_0) \right), \label{An-eq-10} \\
& \text{with }~~~y_p(x_0) = (2p)^{\frac{1}{3}} \left(x_0+\frac{r}{p} \right).\label{An-eq-11}
\end{align}
The task now is to compute the constants $\mathbb{C}_3$ and $\mathbb{C}_4$ for which we use the boundary conditions in Eq. \eqref{An-eq-2} and \eqref{An-eq-3}. Note that $\text{Bi}\left(y_p(x_0) \right)$ diverges as $x_0 \to \infty$ which implies that $\mathbb{C}_4=0$ (see Eq. \eqref{An-eq-3}). The other constant $\mathbb{C}_3$ follows from Eq. \eqref{An-eq-2} as
\begin{align}
\mathbb{C}_3(p,x_R)=\frac{1-\frac{\pi r 2^{1/3}}{p^{2/3}} Q(p, x_R)~\text{Gi}\left(y_p(0) \right)}{\text{Ai}\left(y_p(0) \right)}.
\label{An-eq-12}
\end{align}
Inserting this form of $\mathbb{C}_3(p,x_R)$ in Eq. \eqref{An-eq-10} and setting the initial position $x_0$ equal to the resetting position $x_R$, we get
\begin{align}
&Q(p, x_R) =\frac{\text{Ai}\left(y_p(x_R) \right)}{\text{Ai}\left(y_p(0) \right)\mathcal{M}(p, x_R)},~~~~~~\text{where} \label{An-eq-13}\\
&\mathcal{M}(p, x_R) = 1+ \frac{\pi r 2^{1/3}}{p^{2/3}} \left[ \frac{\text{Ai}\left(y_p(x_R) \right)}{\text{Ai}\left(y_p(0) \right)}\text{Gi}\left(y_p(0) \right)-\text{Gi}\left(y_p(x_R) \right)\right].
\label{An-eq-14}
\end{align} 
For $r=0$, we have $\mathcal{M}(p, x_R)=1$ and $y_p(x_R)  = (2p)^{1/3} x_R$. Substituting these forms in Eq. \eqref{An-eq-13} gives $Q(p, x_R) = \text{Ai}\left((2p)^{1/3} x_R\right) / \text{Ai}(0)$. This matches with the well-known result of Brownian motion in the absence of resetting \cite{Meerson2020, Kearney2005}. 

\begin{figure}[t]
\includegraphics[scale=0.32]{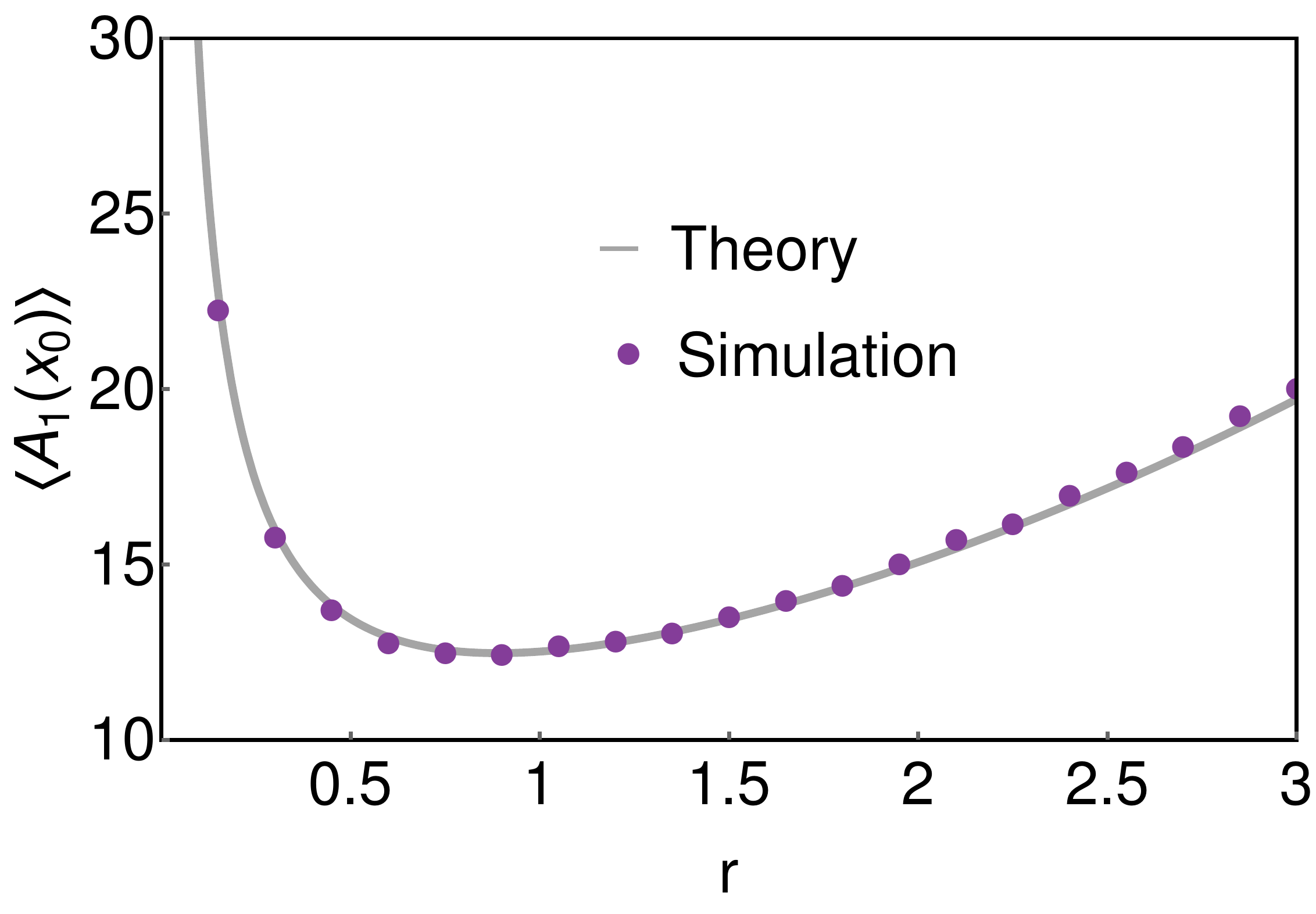}
\includegraphics[scale=0.31]{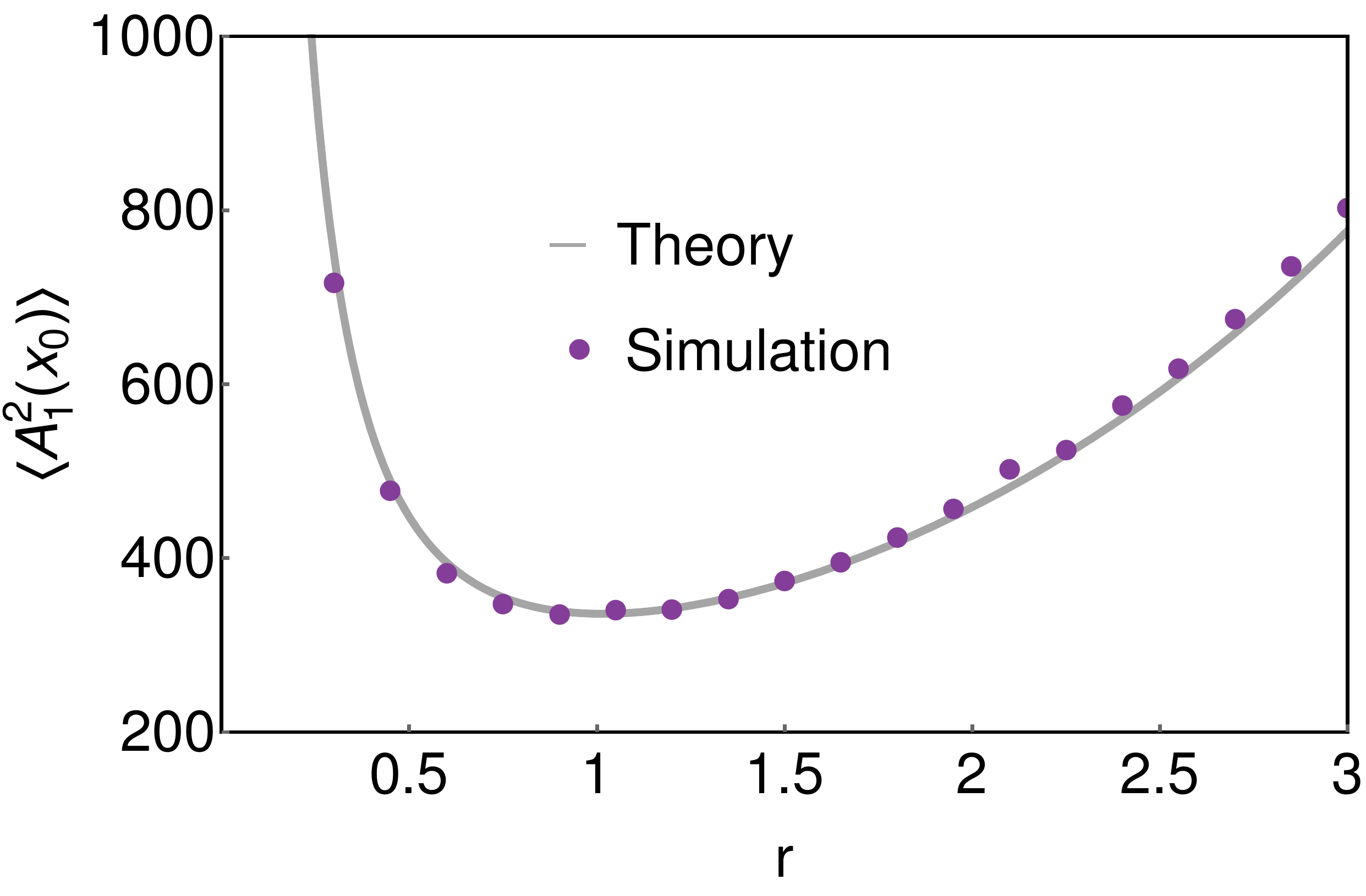}
\centering
\caption{Comparison of the first two moments of the area $A_1(x_0) $ in Eqs. \eqref{An-eq-15} and \eqref{An-eq-16} with numerical simulation. We have taken $x_0=x_R=1.5$ for both plots.}  
\label{area-n-1-moms-Fig}
\end{figure}

Moments of $A_1(x_R)$ can be obtained from $Q(p, x_R)$ using Eq. \eqref{moms-Q}. One can, in principle, compute all the moments of $A_1(x_R)$ from Eq. \eqref{An-eq-13}. Here, we provide expressions for the first two moments:
\begin{align}
\langle A_1(x_R) \rangle & = \frac{x_R}{r}~e^{\sqrt{2r}x_R},     \label{An-eq-15}\\
\langle A_1 ^2(x_R)\rangle & = \frac{2e^{\sqrt{2r}x_R} }{r} \left[ \frac{x_R^2}{r} \left( \frac{3}{4}+e^{\sqrt{2r}x_R}   \right)+\frac{1}{r^2}-\frac{x_R^3}{2\sqrt{2r}}\right] - \frac{2}{r^3}.
\label{An-eq-16}
\end{align} 
In Fig. \ref{area-n-1-moms-Fig}, we have plotted $\langle A_1(x_R) \rangle$ and $\langle A_1 ^2(x_R)\rangle$ as functions of the resetting rate $r$ and compared against the results of simulation. It is worth remarking that the moments of $A_1(x_R)$ for the Brownian motion (without resetting) are infinite as was shown in \cite{Meerson2020}. 
Once again, the moments become finite under the resetting mechanism. Furthermore, we see that the moments exhibit non-monotonic dependence on $r$ with a minimum at some optimal value of $r^*$. We later show that this non-monotonic dependence is a generic feature for all values of $n >-2$.

We now proceed to calculate the distribution of the area $A_1(x_R)$ by performing the inverse Laplace transformation of $Q(p,x_R)$ in Eq. \eqref{An-eq-13}. Performing inverse Laplace transformation for arbitrary $p$ analytically turns out to be difficult. However, one can invert $Q(p,x_R)$ by performing the Bromwich integral numerically. In Fig. \ref{area-dist-Fig} (right panel) we compare thus numerically obtained $P_R(A_1, x_R)$ against simulation data. In order to make some analytic progress, we analyse $Q(p,x_R)$ in limit of small $p$ (equivalently large $A_1$). For small $p$, $y_p(x_R) \to \infty$ as seen from Eq. \eqref{An-eq-11} which in turn implies that the argument of $\text{Ai}\left(y_p(x_R) \right)$ and $\text{Gi}\left(y_p(x_R) \right)$ in Eqs. \eqref{An-eq-13} and \eqref{An-eq-14} become very large. 	We therefore approximate $\text{Ai}\left(y_p \right) \simeq \frac{e^{-\frac{2}{3} y_p^{3/2}}}{2 \sqrt{\pi \sqrt{y_p}}}$ and $\text{Gi}\left(y_p \right) \simeq \frac{1}{\pi y_p}$ for large $y_p$ and insert them in Eq. \eqref{An-eq-13} to get an approximate expression for $Q(p, x_R)$ as
\begin{align}
Q(p,x_R) \simeq \frac{1}{1+p\frac{x_R }{r}e^{\sqrt{2r}x_R}}, ~~~~\text{as }p \to 0.
\label{An-eq-17}
\end{align} 
The inverse Laplace transformation yields an exponential distribution for $P_R\left( A_1,x_R\right)$ namely
\begin{align}
P_R\left( A_1,x_R\right) \simeq \frac{re^{-\sqrt{2r}x_R}}{x_R} ~\text{exp}\left(- \frac{r e^{-\sqrt{2r}x_R}}{x_R} A_1 \right).
\label{An-eq-18}
\end{align}
We emphasise that this expression works only for $A_1 \gg \frac{x_R}{r}~e^{\sqrt{2r}x_R}$. In Fig. \ref{area-dist-Fig} (inset of right panel), we have compared 
this analytical distribution with the same obtained from simulation. We see that there is a mismatch for the small values of $A_1$. However, the prediction becomes accurate as one goes to higher $A_1$. We remark that while the distribution of $A_1$ for Brownian motion (without resetting) has power-law decay of the form $\sim A_1^{-4/3}$ as  $A_1 \to \infty$ \cite{Meerson2020, Kearney2005}, it decays exponentially for $r>0$ demonstrating again the typical behavior of resetting.

\begin{figure}[t]
\includegraphics[scale=0.6]{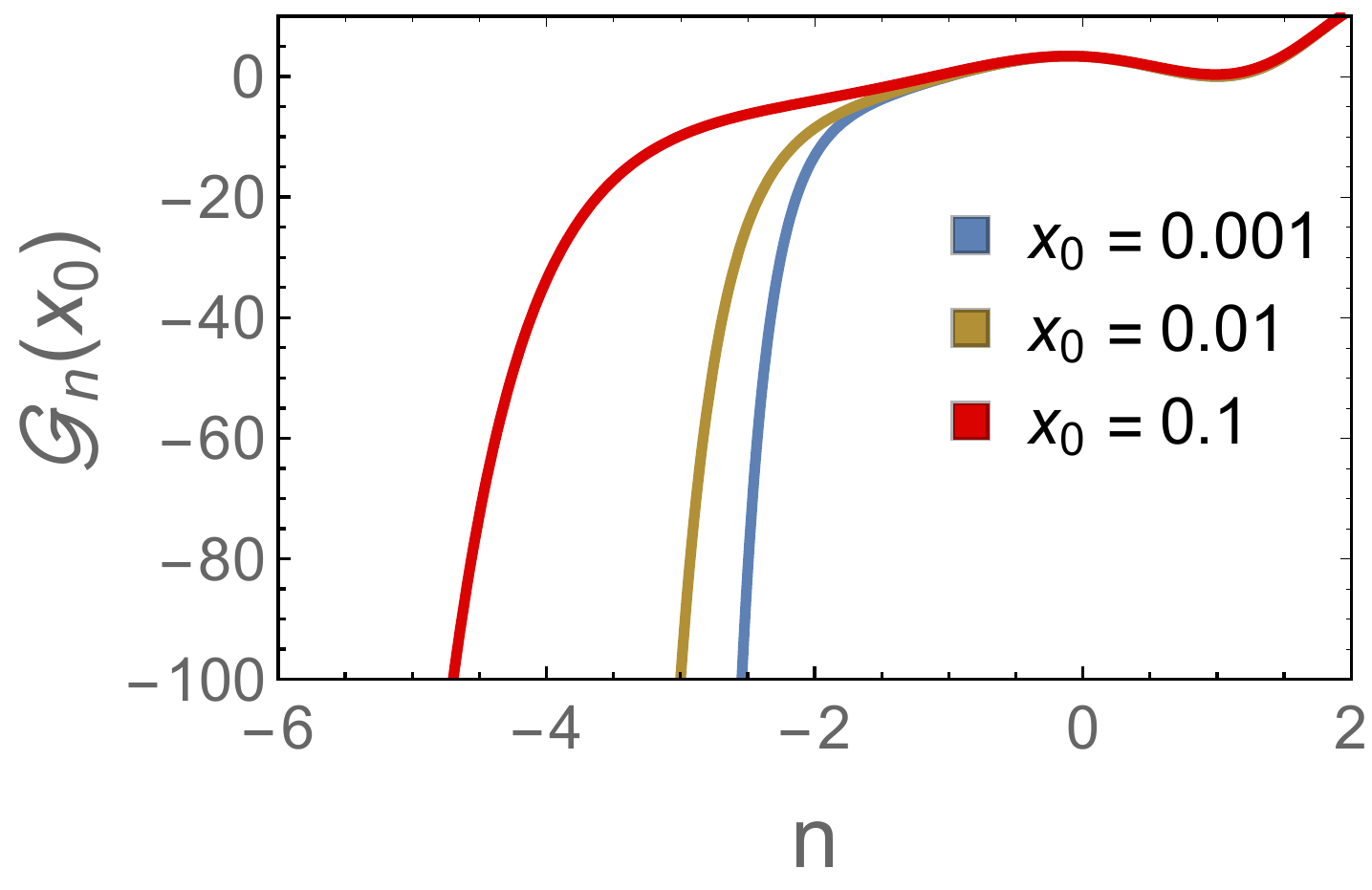}
\centering
\caption{$\mathcal{G}_n(x_0)$ as a function of $n$ for different values of $x_0$ in the limit $x_0 
\to 0$.}   
\label{gn}
\end{figure}

\subsection{General $n~(>-2)$}
We now consider the statistics of $A_n(x_0) = \int _0 ^{t_f} [x(\tau)]^n d\tau$ for general values of $n~(>-2)$. For this, one has to solve the backward Fokker Planck Eq. \eqref{An-eq-1} along with the boundary conditions in Eqs. \eqref{An-eq-2} and \eqref{An-eq-3}. Solving Eq. \eqref{An-eq-1} for arbitrary $n$ and $p$ turns out to be challenging. Here, we present some heuristic analysis that gives exactly the large-$A_n$ behaviour of the distribution $P_R \left(A_n, x_R \right)$.

Numerically, we see that $P_R \left(A_n, x_0 \right)$ decays exponentially for large $A_n$ for all values of $n$. For $n=0$ and $n=1$, we were able to derive this explicitly. Motivated by these observations, we take the following ansatz:
\begin{align}
P_R \left(A_n, x_0 \right) \simeq \frac{e^{-A_n/a_n(x_0)}}{a_n(x_0)},~~~~~~\text{as }\mathcal{A}_n \to \infty,
\label{An-eq-19}
\end{align} 
where $a_n(x_0)$ sets the decay-length for $A_n$. To answer the explicit form of $a_n(x_0)$, we first rewrite Eq. \eqref{An-eq-19} in terms of the Laplace transformation $Q(p,x_0)$ as
\begin{align}
Q(p, x_0) = \frac{1}{1+p~ a_n(x_0)},
\label{An-eq-20}
\end{align}
which is valid only for small $p$ (equivalently for large $A_n$). We next insert this form of $Q(p, x_0)$ in the backward Eq. \eqref{An-eq-1} and up to leading order in $p$, we get
\begin{align}
\frac{1}{2} \frac{d^2a_n}{dx_0^2}  -r a_n(x_0) + r a_n(x_R) =-x_0^n.
\label{An-eq-21}
\end{align}
Solving this equation using \textit{Mathematica}, we find \begin{align}
a_n(x_0) &= a_n(x_R) + \mathbb{C}_5 e^{-\sqrt{2 r} x_0}+\mathbb{C}_6 e^{\sqrt{2 r} x_0}+\mathcal{G}_n(x_0), ~~~\text{with}\label{An-eq-22}\\
\mathcal{G}_n(x_0) &= \text{Re}\left[\frac{e^{\sqrt{2r}x_0} ~\Gamma \left( 1+n, \sqrt{2r}x_0 \right)+(-1)^{-n} e^{-\sqrt{2r}x_0}~ \Gamma \left( 1+n, -\sqrt{2r}x_0 \right)}{(\sqrt{2r})^{n+2}} \right],
\label{An-eq-23}
\end{align}

\begin{figure}[]
\includegraphics[scale=0.205]{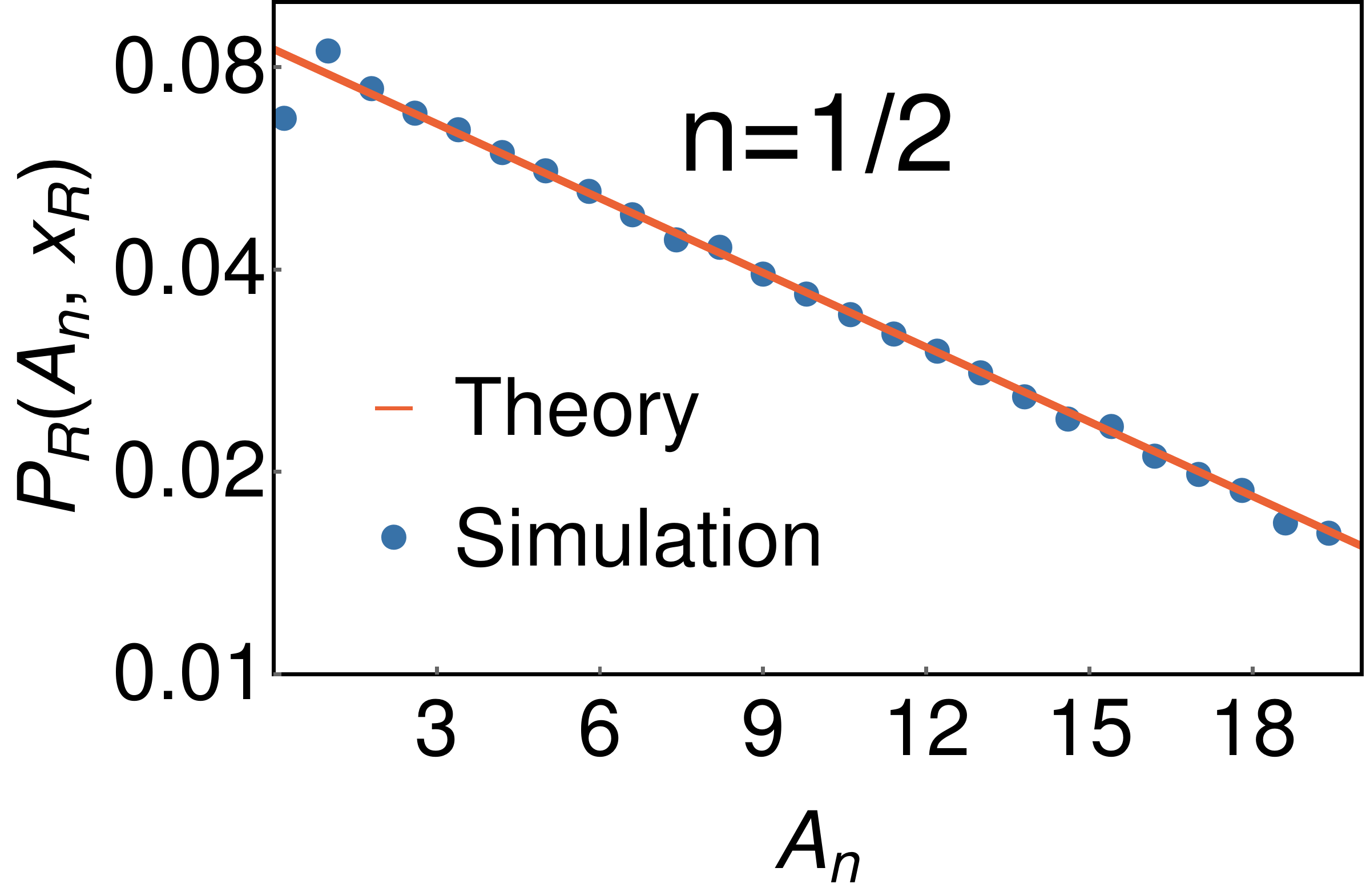}
\includegraphics[scale=0.2]{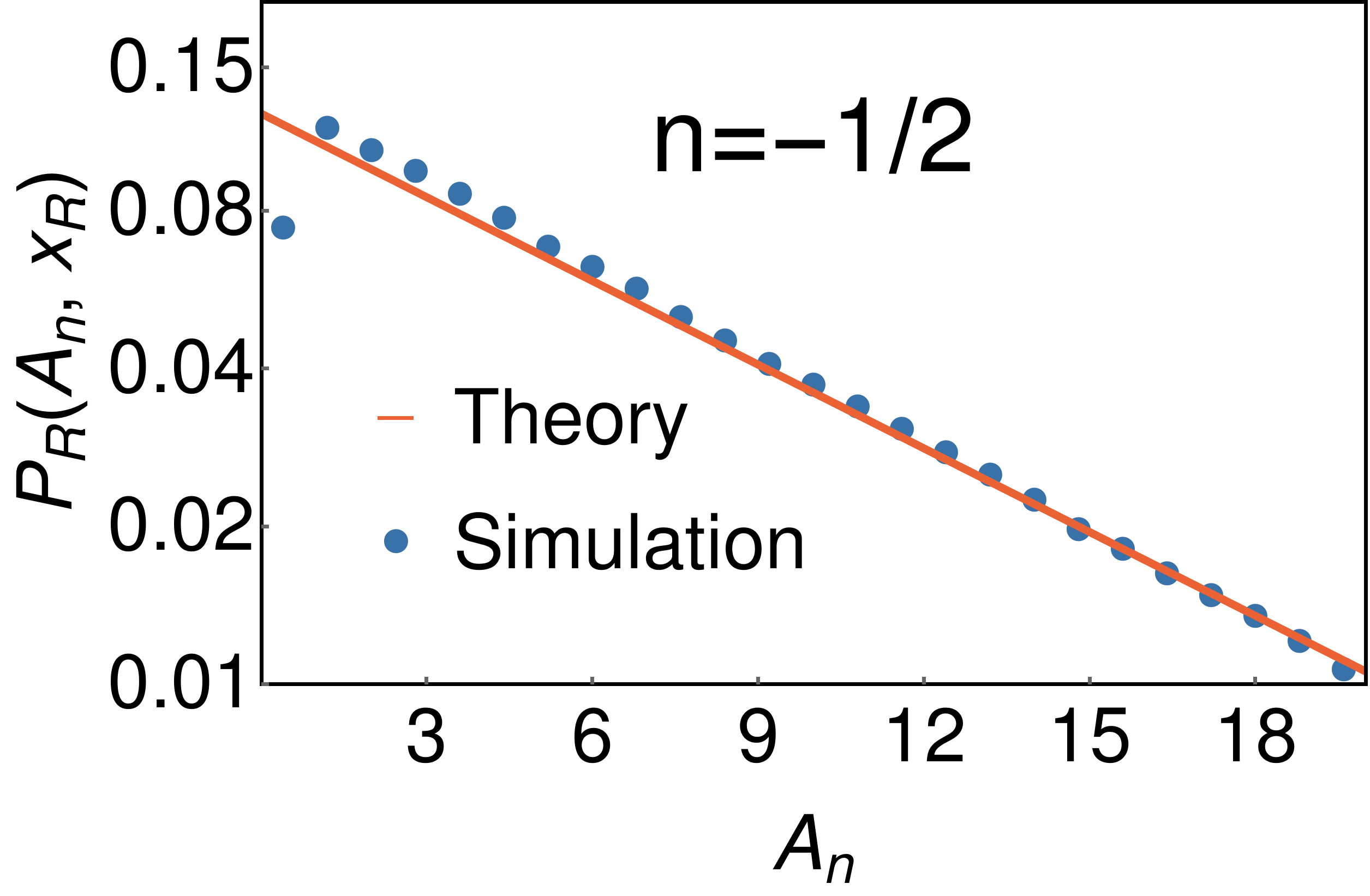}
\includegraphics[scale=0.205]{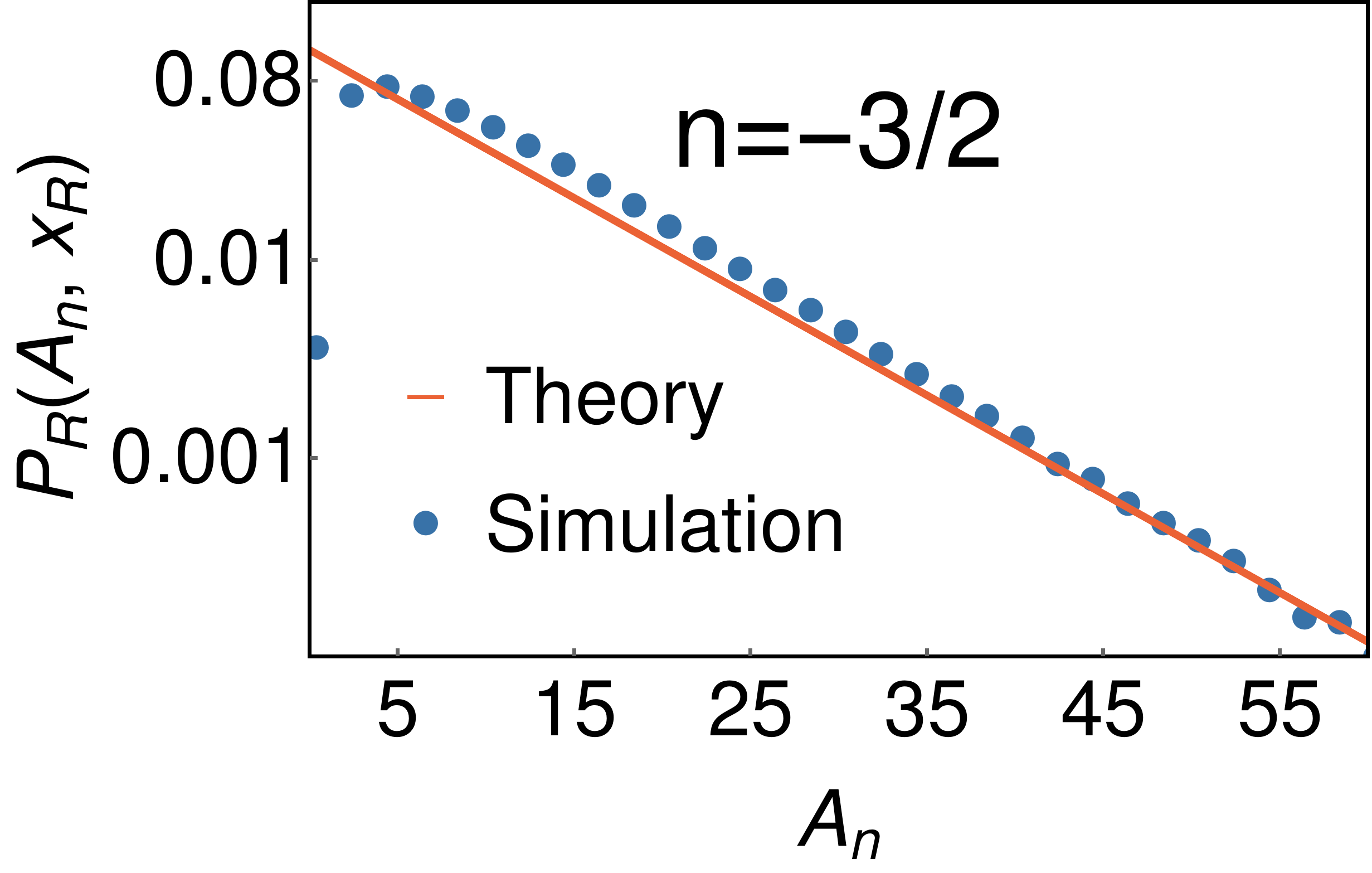}
\centering
\caption{Plot of the distribution $P_R(A_n, x_R)$ in Eq. \eqref{An-eq-19} and comparison with the results of simulation for $n=1/2$ (left), $n=-1/2$ (middle) and $n=-3/2$ (right). Parameters taken are $x_0=x_R=1.5$ and $r=2$.}    
\label{An-dist-Fig}
\end{figure}

\noindent
where $\text{Re}[\cdot]$ indicates the real part of $[\cdot]$. Recall that the function $\mathcal{G}_n(x_0)$ is the particular integral of Eq. \eqref{An-eq-21} and we have taken its real part since the solution $a_n(x_0)$ has to be real. This is also true otherwise since $a_n(x_0)$ represents the mean of $A_n(x_0)$ which is always real.

It is easy to see that the function $\mathcal{G}_n(x_0)$ diverges for $n \leq -2$ as $x_0 \to 0$ (see Fig. \ref{gn}). We later illustrate that this behaviour is not consistent with the boundary condition. In other words, the solution of $a_n(x_0)$ in Eq. \eqref{An-eq-22} is not pertinent for $n \leq -2$. We now proceed to evaluate the constants $ \mathbb{C}_5$ and $ \mathbb{C}_6$ in Eq. \eqref{An-eq-22} for which we use the boundary conditions given by Eqs. \eqref{An-eq-2} and \eqref{An-eq-3}. The former condition gives $a_n(x_0 \to 0) =0$. Using the second condition, we have $a_n(x_0 \to \infty) < \infty$ since the distribution $P_R(A_n, x_0)$ in Eq. \eqref{An-eq-19} is finite for $x_0 \to \infty$. These two conditions evaluate the constants: $ \mathbb{C}_5(p,x_R)=-a_n(x_R)-\mathcal{G}_n(0)$ and $ \mathbb{C}_6(p,x_R)=0$. 
Finally, setting $x_0 = x_R$ we get
\begin{align}
a_n(x_R) = e^{\sqrt{2r}x_R} \mathcal{G}_n(x_R) - \mathcal{G}_n(0).
\label{An-eq-24}
\end{align}  
This expression of $a_n(x_R)$ along with $P _R \left(A_n, x_R\right)$ in Eq. \eqref{An-eq-19} fully characterizes the distribution in the limit of large $A_n$. Our result is consistent with that for the cases $n=0$ and $n=1$ which we rigorously derived. We remark that for simple diffusion without resetting, the distribution of $A_n(x_R)$ exhibits power-law decay i.e., $ P _0 \left(A_n\right) \sim A_n^{-(n+3)/(n+2)} $ for large $A_n$ as was shown by Majumdar and Meerson \cite{Meerson2020}. This behavior remarkably changes as we have shown here that there is a resetting induced exponential decay at least in the asymptotic limit. In Fig. \ref{An-dist-Fig}, we have illustrated this large-$ A_n$ behaviour of $P _R \left(A_n, x_R\right)$ for $n=1/2$ (left panel), $n=-1/2$ (middle panel) and $n=-3/2$ (right panel) and compared with numerical simulations. We observe an excellent agreement between analytical and simulation results for all the cases.

\begin{figure}[]
\includegraphics[scale=0.23]{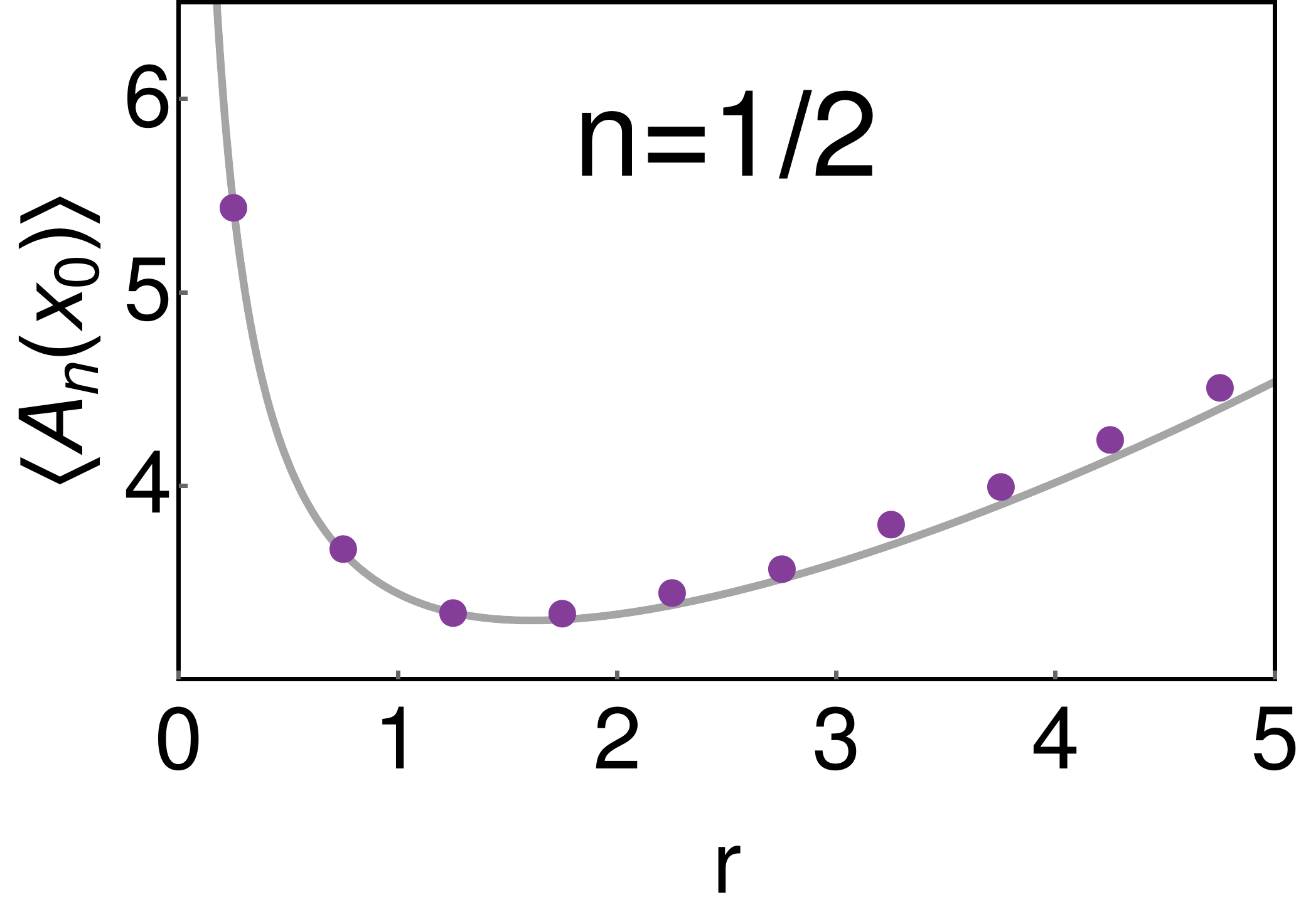}
\includegraphics[scale=0.23]{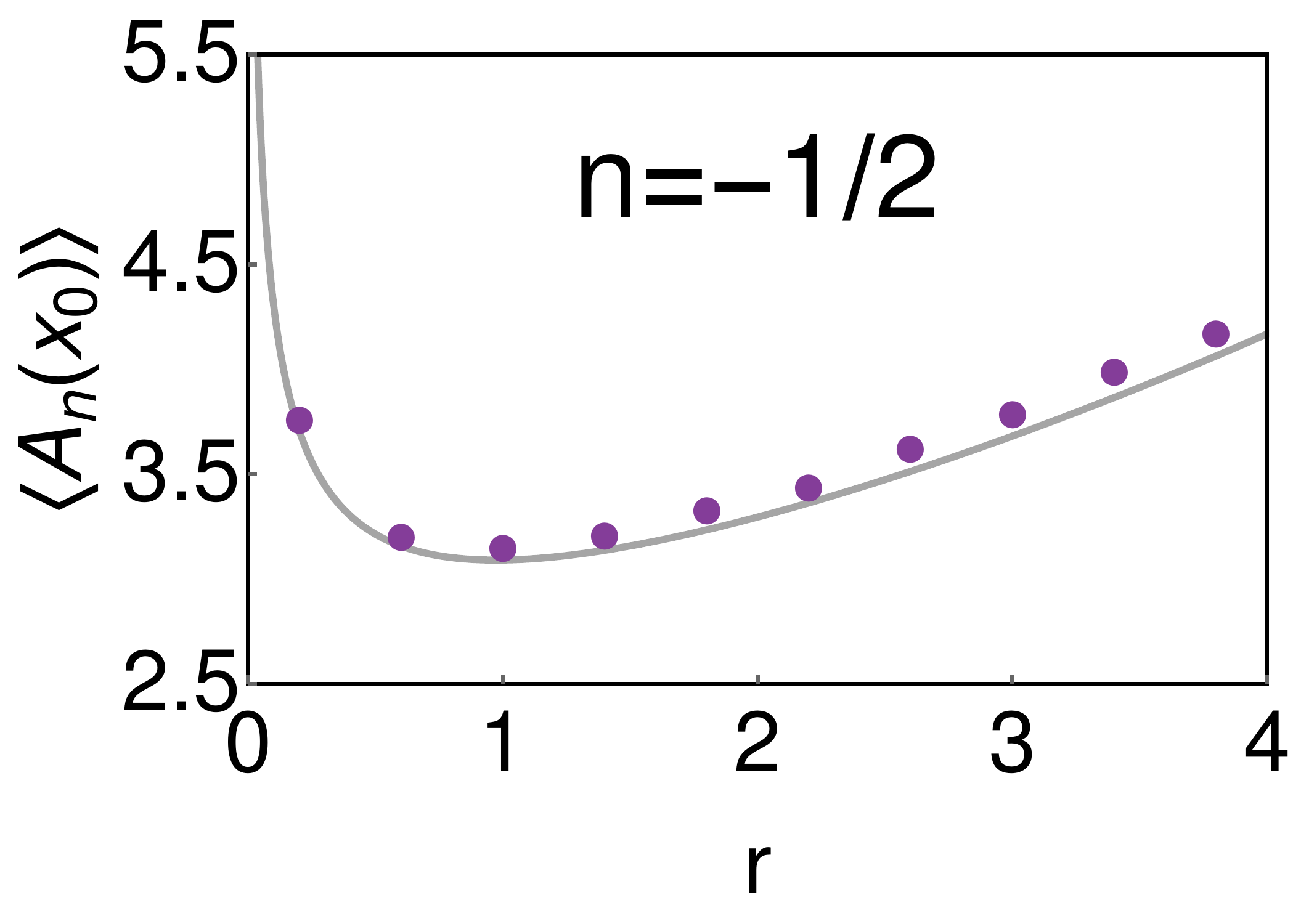}
\includegraphics[scale=0.22]{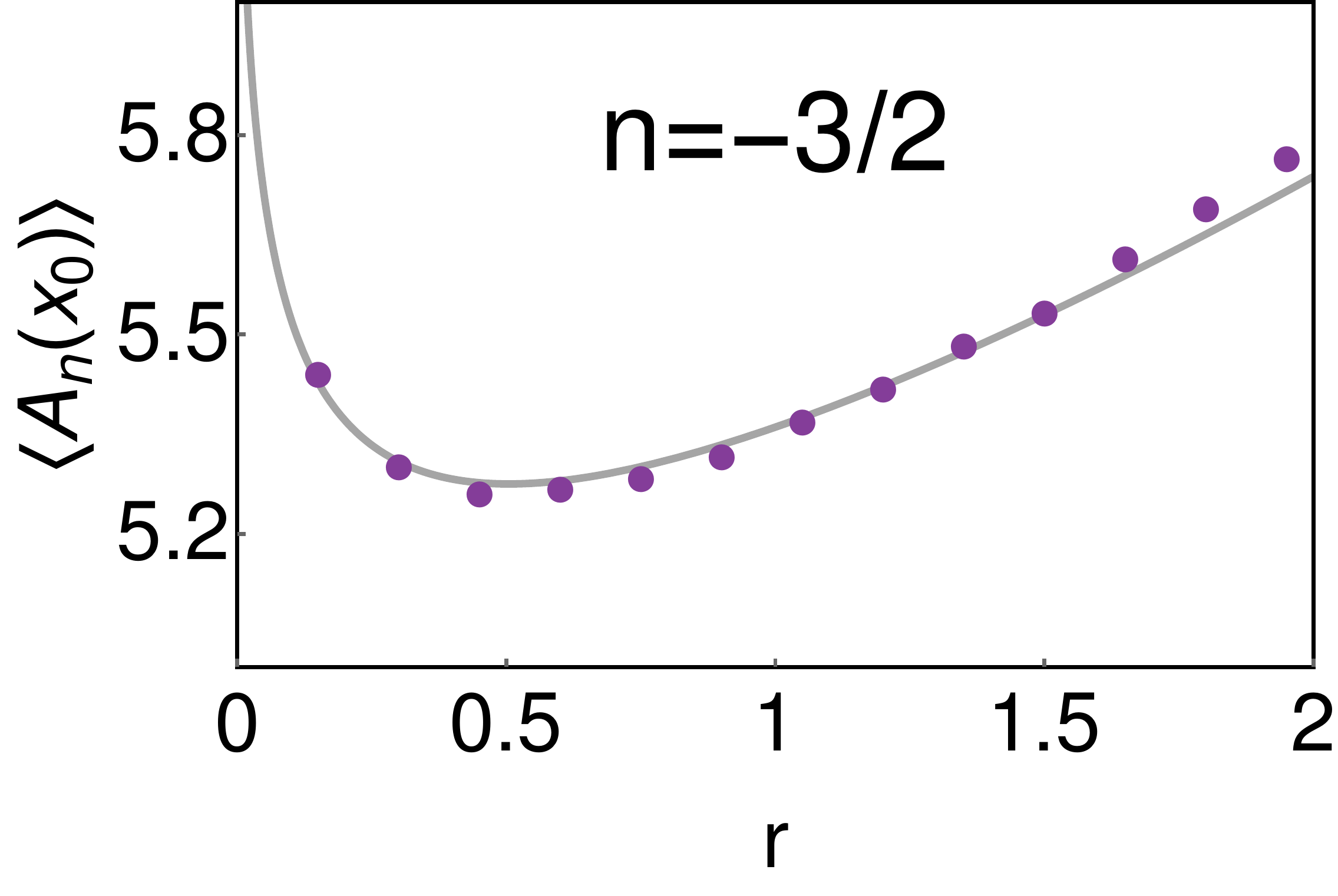}
\centering
\caption{Plots of $\langle A_n(x_R) \rangle$ in Eq. \eqref{An-eq-25} (shown by solid line) for $n=1/2$ (left), $n=-1/2$ (middle) and $n=-3/2$ (right). The theoretical results (solid lines) provide an excellent agreement with the numerical simulations (circles). Parameters in this plot: $x_0 = x_R = 1$.}    
\label{An-mean-Fig}
\end{figure}

Interestingly, the decay length $a_n(x_R)$ in Eq. \eqref{An-eq-24} also turns out to be the mean of $A_n(x_R)$ as stated above. This can be easily verified by rewriting the backward Eq. \eqref{An-eq-1} in terms of $\langle A_n(x_0) \rangle$ which turns out to be same as for $a_n(x_0)$ in Eq. \eqref{An-eq-21}. Therefore, we have
\begin{align}
\langle A_n (x_R)\rangle = e^{\sqrt{2r}x_R} \mathcal{G}_n(x_R) - \mathcal{G}_n(0),
\label{An-eq-25}
\end{align}
where the function $\mathcal{G}_n(x_R)$ is given in Eq. \eqref{An-eq-23}. For $n=0$ and $n=1$, we correctly recover the results of the previous sections. In Fig. \ref{An-mean-Fig}, we have compared the analytical expression of the mean $\langle A_n (x_R)\rangle$ with the results of simulation for $n=1/2$ (left panel), $n=-1/2$ (middle panel) and $n=-3/2$ (right panel). The finiteness of the moments can again be attributed as a direct consequence of resetting. Moreover, the non-monotonic behaviour of the moments is clearly observed as in the case of $n=0$ and $n=1$. 


\section{Conclusion}
\label{conclusion}
In summary, we have characterized the statistical properties of various first passage Brownian functionals in the presence of resetting. Following the famous Feynman-Kac formalism, we presented a general approach that allows for the computation of the local time, residence time, area and other functionals
of a Brownian trajectory that obeys overdamped Langevin dynamics and may also be subject
to stochastic resetting. The process is observed till the first passage time $t_f$ when the particle gets absorbed to the origin for the first time. The central step was to derive a backward differential Eq. \eqref{bfp} for the moment generating function of the functional. This equation was then extensively utilized to derive the exact formulae for the distributions and moments of the local time $(T_{loc})$ and residence time $(T_{res})$. Our study reveals that just like for $t_f$,
the moments of $T_{res}$ also display a non-monotonic dependence on the resetting rate $r$ and exhibit minima at some optimal resetting rate. 

We then turn our attention to the statistics of the functional $A_n(x_0) = \int _0^{t_f} d \tau [x(\tau)]^n$ for $n>-2$. The case of $n=0$ simply represents the first-passage time $t_f$ -- the statistics of which is well understood now. Our central focus was on the non-zero values of $n$. For $n=1$, we again employed the backward Eq. \eqref{bfp} and obtained exact expression of the moment generating function. Utilizing this function, we derive the moments and asymptotic distribution of the area. For this case also, the moments exhibit a non-monotonic dependence on $r$ with minima at some optimal rate $r^*$. Solving Eq. \eqref{bfp} for other values of $n$ turns out to be difficult. Borrowing wisdom from $n=0$ and $n=1$ cases, we presented a heuristic analysis for the general $n$ case which correctly captures the asymptotic form of the distribution of $A_n(x_0)$. For general $n$, we also derived the form of mean $A_n(x_0)$ which, once again, show non-monotonic dependence on $r$. The main conclusion can be summarized in the following way: 
Resetting renders the moments of general functional $A_n(x_0)$ finite for any $n>-2$ similar to the first passage time $t_f$ (which is the $n=0$ case). This is a clear consequence of resetting which removes detrimental realizations that take exceedingly large times to reach the target. Thus, under stochastic resetting, the overall process is completed faster and this feature seems to be quite universal.

Our results can be extended for drift-diffusion processes where drift plays a crucial role in determining the underlying first passage time density. It is known that resetting is detrimental in the high drift limit, while it can expedite the process completion when diffusion dominates over drift. These two phenomena are known to be separated by a resetting transition \cite{RT-1,RT-2,RT-3}. It would be interesting to study these functionals in these two limits and across the transition point. 

Notably, the framework put forward herein is based on the assumption that the resetting is a Markov process. In simple words, resetting time density is exponential and thus memoryless. It would be challenging to extend the formalism under non-exponential resetting time densities. One approach could be to make use of the renewal structure which has been instrumental in connecting observable under resetting with the same in the absence of resetting. However, at this point it not immediately clear how to adapt to that approach for the first passage time functionals. We believe that answering these questions would further deepen our understanding of various first passage time functionals and their behavior in a more general set-up. 

\section{Acknowledgement}
PS acknowledges support from the Department of Atomic Energy, Government of India, under project no.12-R\&D-TFR-5.10-1100. AP acknowledges support from the Department of Atomic Energy, Government of India.

\appendix

\section{Derivation of Eq. \eqref{surv-no-drift-w-reset}}
\label{ILT}
\begin{figure}[]
\includegraphics[scale=0.45]{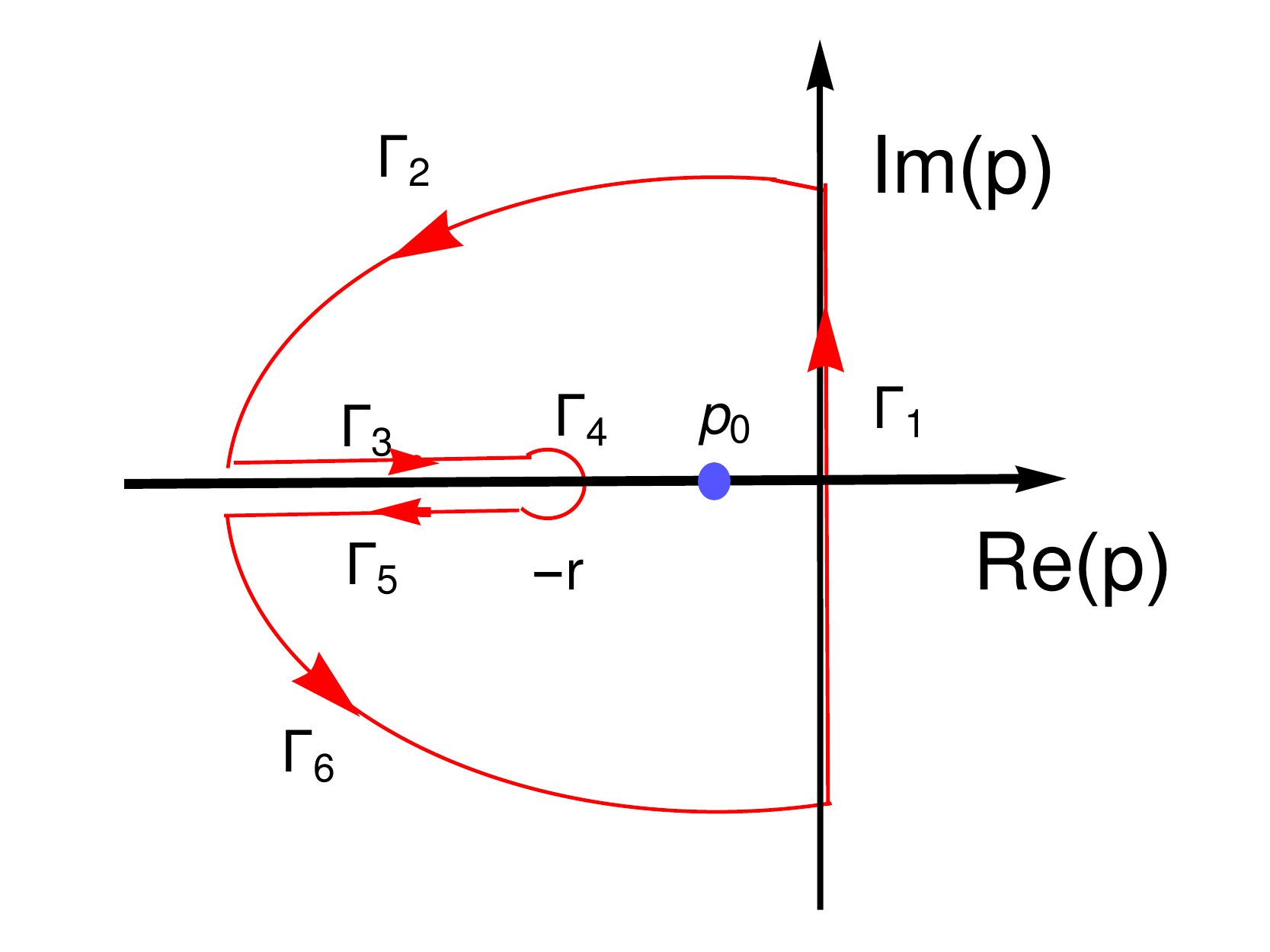}
\centering
\caption{Contour for the Bromwich integral in Eq. \eqref{appen-An-eq--new-1}}    
\label{contour}
\end{figure}

In this appendix, we present the derivation for $P_R(t_f,x_R) $ which was announced in Eq.  \eqref{surv-no-drift-w-reset}. Recall that the distribution in Laplace space is given by (see Eq. \ref{An-eq-5})
\begin{align}
Q(p,x_R) = \frac{(r+p)~e^{-\sqrt{2(r+p)} x_R}}{p+r~e^{-\sqrt{2(r+p)} x_R}}.
\label{appen-An-eq-5}
\end{align}
Formally, the inverse Laplace transform can be written in terms of the Bromwich integral as
\begin{align}
P_R(t_f,x_R) = \frac{1}{2 \pi i}\int _{\gamma _E -i \infty}^{{\gamma _E +i \infty}} dp ~e^{p t_f} Q(p,x_R).
\label{appen-An-eq--new-1}
\end{align}
Here $\gamma _E$ represents a vertical line in the complex plane such that all poles lie to its left (see Fig. \ref{contour}). Plugging $Q(p,x_R)$ from Eq. \eqref{appen-An-eq-5}, we notice that the integral in Eq. \eqref{appen-An-eq--new-1} has a branch point at $p = -r$ and a simple pole at $p_0 =-r(1-u_0)$ with $u_0~(0 \leq u_0 \leq 1)$ obeying the equation $u_0=1-e^{-b_0 \sqrt{u_0}}$ where $b_0 = \sqrt{2 r}x_R$. Note that the pole $p_0$ lies in between $-r$ and $0$. This motivates us to choose $\gamma _E =0$ and consider contour of the form in Fig. \ref{contour} to evaluate Eq. \eqref{appen-An-eq--new-1}.

Since, the integrand has a simple pole inside this contour, Cauchy's theorem states
\begin{align}
\int_{\Gamma _1}+\int_{\Gamma _2}+\int_{\Gamma _3}+\int_{\Gamma _4}+\int_{\Gamma _5}+\int_{\Gamma _6} = \text{Residue of }e^{p t_f} Q(p,x_R) \text{ at }p_0.
\label{appen-An-eq--new-1}
\end{align}
Note that $\int_{\Gamma _1} = P_R(t_f, x_R)$ is the required integral. We first compute the residue at $p = p_0$.
\begin{align}
\text{Residue at }p_0 & = \lim _{p \to p_0} \left[ (p-p_0)e^{p t_f} Q(p,x_R) \right] \\
& = \left[\frac{r u_0 ~e^{-b_0 \sqrt{u_0}}}{1-\frac{b_0}{2 \sqrt{u_0}} e^{- b_0 \sqrt{u_0}}} \right]e^{-r(1-u_0) t_f}.
\label{appen-An-eq--new-2}
\end{align}
Next we proceed to compute the contribution of integrals across different paths in Eq. \eqref{appen-An-eq--new-1}. Recall that the real part of $p$ along $\Gamma _2$ and $\Gamma _6$ is negative and in the limit $|p| \to \infty$, this contribution becomes zero. On the other hand, for $\Gamma _4$, we substitute $p = - r +\epsilon e^{i \theta}$ and take $\epsilon \to 0^+$ limit. The integrand, in this limit, becomes exactly zero, i.e. $\int _{\Gamma _4}=0$. Hence, we only have to evaluate integrals along $\Gamma_3$ and $\Gamma _5$. 
To evaluate $\int _{\Gamma _3}$, we substitute $p=-r+w e^{i \pi}$ to obtain
\begin{align}
\int _{\Gamma _3} =\frac{1}{2 \pi i} \int _0^{\infty} dw~e^{-(r+w)t_f} \frac{w e^{-i \sqrt{2w} x_R}}{r+w-re^{-i \sqrt{2w} x_R}}.
\label{appen-An-eq--new-3}
\end{align}
Similarly, for $\int _{\Gamma _5}$, we substitute $p=-r+w e^{-i \pi}$ to get
\begin{align}
\int _{\Gamma _5} =-\frac{1}{2 \pi i} \int _0^{\infty} dw~e^{-(r+w)t_f} \frac{w e^{i \sqrt{2w} x_R}}{r+w-re^{i \sqrt{2w} x_R}}.
\label{appen-An-eq--new-4}
\end{align}
Inserting all these contributions in Eq. \eqref{appen-An-eq--new-1} and performing some algebraic simplications, we find
\begin{align}
P_R(t_f,x_R) & = \left[\frac{r u_0 ~e^{-b_0 \sqrt{u_0}}}{1-\frac{b_0}{2 \sqrt{u_0}} e^{- b_0 \sqrt{u_0}}} \right]e^{-r(1-u_0) t_f}+ H_r(t_f), ~~~~~~~~\text{with}\label{appen-An-eq--new-5}\\
 H_r(t_f) & = \frac{e^{-r t_f}}{\pi} \int _{0}^{\infty} dw ~e^{-w t_f}~\frac{w(r+w) \sin \left(\sqrt{2 w} x_R \right)}{(r+w)^2+r^2-2 r(r+w) \cos \left(\sqrt{2 w} x_R \right)}.
\end{align}
This result has been quoted in Eq. \eqref{surv-no-drift-w-reset}

\end{document}